\documentclass[epsfig,12pt]{article}
\usepackage{makeidx}
\usepackage{amsmath}
\usepackage{amsfonts}
\usepackage{amssymb}
\usepackage{graphicx}
\usepackage{accents}

\input epsf.sty
\newtheorem{theorem}{Theorem}
\newtheorem{acknowledgement}[theorem]{Acknowledgement}

\makeatletter \@addtoreset{equation}{section}
\renewcommand{\theequation}{\thesection.\arabic{equation}}
\input{tcilatex}
\begin{document}

\title{\rightline{\mbox{\small
{Lab/UFR-HEP0701/GNPHE/0701}}} \textbf{Non Planar Topological 3-Vertex
Formalism}\\
}
\author{Lalla Btissam Drissi$^{1,2}${\small \thanks{%
drissilb@gmail.com}} , Houda Jehjouh$^{1,2}${\small \thanks{%
jehjouh@gmail.com}} , El Hassan\ Saidi$^{1,2,3}${\small \thanks{%
h-saidi@fsr.ac.ma}} \\
{\small 1.} {\small Lab/UFR- Physique des Hautes Energies, Facult\'{e} des
Sciences, Rabat, Morocco,}\\
{\small 2.} {\small GNPHE, Groupement National de Physique des Hautes
Energies, }\\
{\small Si\`{e}ge focal: Facult\'{e} des Sciences, Rabat, Morocco}\\
{\small 3. Coll\`{e}ge SPC, Acad\'{e}mie Hassan II des Sciences et
Techniques, Rabat, Morocco}} \maketitle

\begin{abstract}
Using embedding of complex curves in the complex projective plane $\mathbb{P%
}^{2}$, we develop a \emph{non planar} topological 3-vertex formalism for
topological strings on the family of local Calabi-Yau threefolds $X^{\left(
m,-m,0\right) }=\mathcal{O}(m)\oplus \mathcal{O}(-m)\rightarrow E^{\left(
t,\infty \right) }$. The base $E^{\left( t,\infty \right) }$ stands for the
degenerate elliptic curve with Kahler parameter $t$; but a large complex
structure $\mu $; i.e $\left\vert \mu \right\vert \longrightarrow \infty $.
We also give first results regarding A-model topological string amplitudes
on $X^{\left( m,-m,0\right) }$. The $2D$ $U\left( 1\right) $ gauged $%
\mathcal{N}=2$ supersymmetric sigma models of the degenerate elliptic curve $%
E^{\left( t,\infty \right) }$ as well as for the family $X^{\left(
m,-m,0\right) }$ are studied and the role of D- and F-terms is explicitly
exhibited. \newline
\end{abstract}


\section{ Introduction}

\qquad Topological string theory \cite{01,02,03} is a powerful method to
deal with the $4D$ $\mathcal{N}=2$ supergravity Planck limit of the
compactification of type II superstring on Calabi-Yau (CY) threefolds X$_{3}$
\cite{1,2}. The study of \emph{local }Gromov-Witten theory of curves in
\emph{non compact} Calabi-Yau threefolds \cite{201,21,G1,h,G2} and the OSV
conjecture \cite{3}\textrm{,} relating microscopic \emph{4D} black holes to
\emph{2D} q-deformed Yang-Mills theory\textrm{\ }\cite{4}-\cite{8}, have
given an additional impulse to the revival interest in the study of the
topological field and string theories \cite{80}-\cite{85}.\ Several
important results have been obtained in the few last years; in particular
the development of the topological tri-vertex ( to which we refer below as
\emph{planar} 3-vertex) method for computing the A-model partition function
for non compact toric CY3s \cite{9,10} and the interpretation of this vertex
in terms of 3d-partitions of melting crystals generalizing $U\left( \infty
\right) $ Young tableau \textrm{\ }\cite{11,12}.

\qquad Moreover the \emph{planar} 3-vertex and its refined version \cite%
{12a,12b} have been shown particularly powerful. They agree with the
Nekrasov's partition function of $\mathcal{N}=2$ $SU(N)$ gauge theory $\cite%
{120}$-$\cite{124}$ and provide more insights into non perturbative dynamics
of string field theory. The power of the topological 3-vertex method may be
compared with the power of Feynman graphs technique in perturbative $\phi
^{3}$ quantum field theory (QFT). This formal similarity between the toric
web-diagrams and the Feynman graphs opens a window on the following issues:%
\newline
\textit{First}, the use of perturbative QFT results to motivate topological
stringy analogues, in particular toric web-diagrams with higher dimensional
vertices such as the typical $\phi ^{4}$ to be considered in this study;
\newline
\textit{Second}, the development of new techniques to enlarge the class of
toric Calabi-Yau \textrm{threefolds} to which the topological vertex
formalism applies.

\qquad Recall that for \emph{non compact toric} Calabi-Yau threefolds $X_{3}$
with toric web-diagram $\Delta \left( X_{3}\right) $, the \emph{planar}
3-vertex method allows to compute explicitly the A- model topological string
amplitudes. The topological vertex method is a Feynman-rules like technique
where the Feynman graphs, the vertices of these graphs, the momenta, and the
propagators correspond respectively to the toric web-diagrams $\Delta \left(
X_{3}\right) $, the 3-valent vertices $C_{\lambda \mu \nu }$, Young diagrams
$\lambda $, and the weights $\left( -\right) ^{\left( n+1\right) \left\vert
\lambda \right\vert }e^{-t\left\vert \lambda \right\vert }q^{-\frac{n}{2}%
\kappa \left( \lambda \right) }$ where $n$ encodes the framing.

\qquad Motivated by: \newline
(\textbf{1}) the \emph{formal correspondence} between toric web-diagrams of
local Calabi-Yau threefolds and QFT Feynman graphs, \newline
(\textbf{2}) the two classes of toric Calabi-Yau threefolds describing the
vacua of supersymmetric sigma model with (W$\left( \Phi _{i}\right) \neq 0$)
and without superpotential (W$\left( \Phi _{i}\right) =0$) , and \newline
(\textbf{3}) a special feature\textrm{\footnote{%
In toric Calabi-Yau 3-folds $X_{3}$ with typical fibration $B\times F$, the
torii appear generally in the fiber $F$. In the local elliptic curve we are
considering in this paper, the 2-torus is in the base $B$.}} of the
{\normalsize local 2- torus} $\mathcal{O}(m)\oplus \mathcal{O}%
(-m)\rightarrow E^{\left( t,\infty \right) }$ where the elliptic curve%
\textrm{\footnote{%
the 2-torus has one Kahler parameter $t$ and one complex parameter $\mu $.
As these parameters play an important role here, we will exhibit them below
by referring to the elliptic curve as $E^{\left( t,\mu \right) }$. Further
details are given in the appendix.}}
\begin{equation}
E^{\left( t,\mu \right) }
\end{equation}%
is in the base of the local Calabi-Yau threefold $X^{\left( m,-m,0\right) }$
rather than in the fiber, \newline
we address in this paper, the two following points: \newline
(\textbf{a}) We propose in this study a toric representation for the family
of the local 2-torii with{\normalsize \ }fixed finite Kahler parameter $t$
and a \textit{large complex structure }$\mu $\textit{; say }$\left\vert \mu
\right\vert \longrightarrow \infty $\textit{,}
\begin{equation}
\mathcal{O}(m)\oplus \mathcal{O}(-m)\rightarrow E^{\left( t,\infty \right) },
\label{P1}
\end{equation}%
with integer $m$. The \textit{degenerate} elliptic curve
\begin{equation}
E^{\left( t,\infty \right) },
\end{equation}%
describing the base of the above local Calabi-Yau threefolds (\ref{P1}),
will be realized as the (toric) boundary of the complex toric projective
plane $\mathbb{P}^{2}$; see sections 3, 5 and 6 as well as the appendix for
more details. \newline
With this representation at hand, we can then:\newline
($\mathbf{\alpha }$) circumvent, at least for the particular case of the
degenerate $E^{\left( t,\infty \right) }$, the usual difficulty regarding
the lack of a toric diagram for the 2-torus. \newline
In addition to the large complex structure limit constraint $\left\vert \mu
\right\vert \rightarrow \infty $, the other price to pay in this set up is
to consider a \emph{non planar} 3-vertex formalism rather than the standard
\emph{planar} 3-vertex one based on the $R\times T^{2}$ special Lagrangian
fibration of $\mathbb{C}^{3}$. The reason behind the emergence of the \emph{%
non planar} 3-vertex is that the toric Calabi-Yau 3-fold $X^{\left(
m,-m,0\right) }$ is realized as a non compact toric hypersurface of the
complex four dimensional toric manifold
\begin{equation}
\mathcal{O}(m)\oplus \mathcal{O}(-m)\rightarrow \mathbb{P}^{2}.  \label{P2}
\end{equation}%
For later use, we will refer to the local geometries (\ref{P1}) and (\ref{P2}%
) as $H_{3}$ and $Y_{4}$ respectively. \newline
The geometry $Y_{4}$ will be also promoted to the Calab-Yau 4-fold $X_{4}=%
\mathcal{O}(-m-3)\rightarrow W\mathbb{P}_{1,1,1,m}^{3}$.\newline
($\mathbf{\beta }$) draw the lines for computing the topological amplitudes
by using a \emph{non planar} 3-vertex formalism.\newline
In the present study, we will mainly set up the key idea by:\newline
($\emph{i}$) building the toric web-diagram $\Delta \left( X_{3}\right) $ of
the local degenerate elliptic curve $X^{\left( m,-m,0\right) }$.\newline
($\emph{ii}$) give first results regarding the structure of the topological
\emph{non planar} 3-vertex and the partition function of $X^{\left(
m,-m,0\right) }$ as well as their relation to the 4- vertex of the ambient $%
\mathbb{C}^{4}$ local patch and to the generalized Young diagrams.\newline
(\textbf{b}) We develop the supersymmetric linear sigma model field theory
setting of the local degenerate elliptic curve $X^{\left( m,-m,0\right) }$.
\newline
More precisely, we show the two main following things:\newline
($\mathbf{\alpha }$) the \emph{planar} 3-vertex method is associated with
the auxiliary \emph{D-terms} in supersymmetric sigma models. \newline
The \emph{non planar} 3-vertex formalism we will be considering here
corresponds to the case where we have both \emph{D-terms} and \emph{F-terms}%
. \newline
($\mathbf{\beta }$) We use the sigma model for local $\mathbb{P}^{2}$ to
induce the $\mathcal{N}=2$ supersymmetric gauged model for the local
elliptic curve $X^{\left( m,-m,0\right) }$. The underlying complex geometry
of a such construction was noticed in the Witten's original work\ \cite{18}.
Here, we give explicit details regarding the implementation of \emph{F-terms}%
.

\qquad The organization of this paper is as follows: In section 2, we give
an overview of the topological \emph{planar} 3- vertex method. In section 3,
we exhibit briefly first results concerning the topological \emph{non planar}
3-vertex by considering the example of a Calabi-Yau threefold $H_{3}$. The
toric 3-fold $H_{3}$ is realized as a hypersurface of a four dimensional
complex Kahler manifold.\emph{\ }In section 4, we review the main points of
the $U\left( 1\right) $ gauged supersymmetric sigma model realization of the
local $\mathbb{P}^{2}$. In section 5, we consider the sigma model for the
degenerate local elliptic curve $X^{\left( m,-m,0\right) }$. As the question
of the toric realization of $\mathbb{T}^{2}$ is a crucial point, we divide
this section in three parts: We first study the realization of the
degenerate elliptic curve $E^{\left( t,\infty \right) }$ by using the
compact divisor of $\mathbb{P}^{2}$. Then we give explicit details regarding
the $U\left( 1\right) $ gauged supersymmetric sigma\ model realization of
the local degenerate elliptic curve $X^{\left( m,-m,0\right) }$. Next, we
study the moduli space of the supersymmetric vacua associated with $%
X^{\left( 3,-3,0\right) }$. In section 6, we extend the construction to the
case of local elliptic curve $X^{\left( m,-m,0\right) }$. In section 7, we
give the conclusion and in section 8 we give an appendix where we show that $%
\partial \left( \mathbb{P}^{2}\right) $ is precisely $E^{\left( t,\infty
\right) }$.

\section{Topological vertex method}

\qquad In this section, we consider the topological 3-vertex method used for
the computation of the A- model topological string amplitudes. We illustrate
this method through some examples of \emph{non compact} toric Calabi-Yau
threefolds namely: \newline
(\textbf{1}) the complex space $\mathbb{C}^{3}$, with special Lagrangian
fibration as $R\times T^{2}$, playing the role of the \emph{planar} 3-vertex.%
\newline
(\textbf{2}) the resolved conifold obtained by gluing two \emph{planar}
3-vertices,\newline
(\textbf{3}) local $\mathbb{P}^{2}$ made of three \emph{planar} 3-vertices.%
\newline
Then, we consider an example of toric Calabi-Yau threefold (\ref{P1}) where
one needs introducing \emph{non} \emph{planar }3-vertices\emph{\ }(and
4-vertices). This local Calabi-Yau threefold is precisely the one given by
the local degenerate elliptic curve $X^{\left( m,-m,0\right) }=\mathcal{O}%
(m)\oplus \mathcal{O}(-m)\rightarrow E^{\left( t,\infty \right) }$ realized
as a hypersurface in (\ref{P2}).

\subsection{Tri-vertex method: A brief review}

\qquad The topological 3-vertex formalism computes the partition function
\begin{equation*}
Z_{X_{3}}(q)
\end{equation*}%
of the local toric Calabi-Yau threefolds $X_{3}$. In this formalism, the
toric web-diagram of $X_{3}$ is thought of as resulting from gluing copies
of \emph{planar} 3-vertices $C_{\lambda \mu \nu }$ along their edges.
\newline
Recall that the topological vertex $C_{\lambda \mu \nu }$ has three legs
ending on stacks of Lagrangian D-branes ($\mathbb{C}\times S^{1}$)
represented by 2d partitions $\lambda $, $\mu $ and $\nu $.\newline
The partition function $Z_{X_{3}}(q)$ depends on the following quantities:%
\newline
(\textbf{i}) the parameter $q$ which reads in terms of the string coupling
as $e^{-g_{s}}$; it plays the role of the Boltzmann weight.\newline
(\textbf{ii}) the Kahler parameters $\left\{ t_{i}\right\} $ of the local
Calabi-Yau threefold $X_{3}$ ( $i=1,...,h^{1,1}\left( X_{3}\right) $).
\newline
Below, we will consider simple examples where
\begin{equation*}
h^{1,1}\left( X_{3}\right) =1.
\end{equation*}%
(\textbf{iii}) the boundary conditions (open strings) described by $2d$
partitions $\mu $ ( generic representations of $U\left( \infty \right) $).
In the QFT language where Feynman graphs play a quite similar role as the
toric web-diagrams, the 2d partition $\mu $ corresponds to\emph{\ }the\emph{%
\ "external momentum" }of Feynman graph. Recall that a 2d partition $\mu $
is a Young diagram with columns
\begin{equation}
\left( \mu _{1},\mu _{2},...\right) ,\qquad \mu _{i}\geq \mu _{i+1},\qquad
\mu _{i}\in \mathbb{Z}_{+}.  \label{2p}
\end{equation}%
Columns of the 2d partition are associated with Lagrangian D- branes and
rows with Lagrangian anti- D-branes.\newline
(\textbf{iv}) Lagrangian D-brane/anti-D-brane pairs are needed for the
gluing of the vertices. The gluing operation is achieved by inserting 2d
partitions $\nu $ and their transpose $\nu ^{T}$ at the cuts and summing
over all possible $\nu $'s. In\emph{\ }QFT language,\emph{\ }$\nu $\emph{\ }%
corresponds\emph{\ to "internal momenta}".\newline
The topological 3-vertex\footnote{%
For simplicity, we use 3-vertex to refer to the planar 3-vertex.} method for
computing the partition function $Z_{X_{3}}(q)$ is illustrated on the three
examples given below.

\subsection{Examples}

\emph{Example 1}: the 3-vertex of $\mathbb{C}^{3}$\newline
The toric graph of $\mathbb{C}^{3}$ is given by \emph{figure 1}.
\begin{figure}[tbph]
\begin{center}
\hspace{-1cm} \includegraphics[width=5cm]{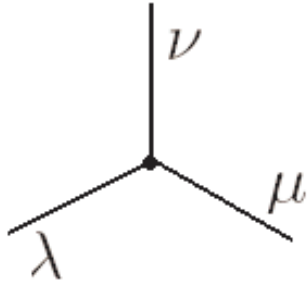}
\end{center}
\par
\vspace{-0.5cm}
\caption{{\protect\small The toric web-diagram of }$C^{3}${\protect\small .
It appears as local patches in toric Calabi-Yau threefolds. The three edges
end on stacks of Lagrangian D branes. }$\protect\lambda ,${\protect\small \ }%
$\protect\mu ${\protect\small \ and }$\protect\nu ${\protect\small \ are 2d
partitions which, in QFT set up, may be thought of as the external momenta}.}
\label{3v}
\end{figure}
Following \cite{9}, the partition function of the 3-vertex, with a stack of%
\emph{\ }Lagrangian D-branes ending on its legs captured by the boundary
conditions\emph{\ }$\left( \lambda ,\mu ,\nu \right) $, is given by
\begin{equation}
Z_{X_{3}}(q)=\sum_{\lambda ,\mu ,\nu }C_{\lambda \mu \nu }(q)\left(
Tr_{\lambda }\mathcal{V}\text{ }Tr_{\mu }\mathcal{V}\text{ }Tr_{\nu }%
\mathcal{V}\right) .
\end{equation}%
In this relation, the trace $Tr_{\lambda }$ of the holonomy matrix $\mathcal{%
V}$ , with eigenvalues $x=\left( x_{1},x_{2},...\right) $, is given by the
Schur function $\mathcal{S}_{\lambda }(x)$. The latter depends on the 2d
partition $\lambda =\left( \lambda _{1},\lambda _{2},...\right) $ and the $%
x_{i}=q^{i-1/2-\lambda _{i}}$. The rank three tensor
\begin{equation}
C^{\left( 3\right) }=C_{\lambda \mu \nu },  \label{c3}
\end{equation}%
is the topological 3-vertex whose explicit expression reads as
\begin{equation}
C_{\lambda \mu \nu }(q)=q^{\kappa (\lambda )}\left[ \mathcal{S}_{\nu
^{T}}(q^{-\rho })\sum_{2d\text{ partitions }\eta }\mathcal{S}_{\lambda
^{T}/\eta }(q^{-\nu -\rho })\mathcal{S}_{\mu /\eta }(q^{-\nu ^{T}-\rho })%
\right]  \label{c}
\end{equation}%
with $\rho =\left( \rho _{1},\rho _{2},...\right) $ and $\rho _{i}=1/2-i$.
\newline
Eq(\ref{c}) involves the product of skew-Schur functions $\mathcal{S}_{\mu
/\eta }$. It reduces, for the closed topological string case, to
\begin{equation}
Z_{\mathbb{C}^{3}}(q)=C_{\emptyset \emptyset \emptyset
}(q)=\prod\limits_{n=1}^{\infty }(1-q^{n})^{-n}.
\end{equation}%
which is nothing but the 3d MacMahon function.\newline

\emph{Example 2}: \qquad Resolved conifold $\ X_{3}=\mathcal{O}\left(
-1\right) \oplus \mathcal{O}\left( -1\right) \rightarrow \mathbb{P}^{1}$%
\newline
The resolved conifold has one Kahler parameter $t$ parameterizing the size
of the projective line $\mathbb{P}^{1}$. \emph{Figure 2} describes its toric
web-diagram.\newline
\begin{figure}[tbph]
\begin{center}
\hspace{-1cm} \includegraphics[width=6cm]{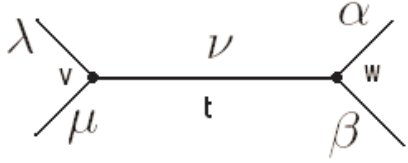}
\end{center}
\par
\vspace{-1cm}
\caption{ Resolved conifold $\mathcal{O}\left( -1\right) \oplus \mathcal{O}%
\left( -1\right) \rightarrow P^{1}$ made of two planar 3-vertices.}
\label{23v}
\end{figure}
\newline
This local threefold $X_{3}$ is obtained by gluing \emph{two} 3-vertices
along one edge leaving then four opened external legs. \newline
In the simplest case where there is no boundary terms on the external legs,
the partition function of the closed topological string on the resolved
conifold is given by,
\begin{equation}
Z_{X_{3}}(q,t)=\sum_{\text{2d partitions }\nu }C_{\emptyset \emptyset \nu
}(q)\text{ }(-1)^{\left\vert \nu \right\vert }e^{-\left\vert \nu \right\vert
t}\text{ }C_{\emptyset \emptyset \nu ^{T}}(q)\text{.}
\end{equation}%
In this relation, $\nu ^{\mathrm{T}}$ is the transpose of the 2d partition $%
\nu $ with $\left\vert \nu \right\vert $ boxes and $C_{\emptyset \emptyset
\nu }(q)$ is as in eq(\ref{c}) by setting the boundary conditions as $%
\lambda =\emptyset $ and $\mu =\emptyset $.\newline

\emph{Example 3}: \qquad \emph{Local} $\mathbb{P}^{2}:$ $\ X_{3}=\mathcal{O}%
\left( -3\right) $ $\ \mathbb{\rightarrow P}^{2}$\newline
The local $\mathbb{P}^{2}$ has one Kahler modulus $t$ parameterizing the
size of the projective plane $\mathbb{P}^{2}$. \emph{Figure 3} describes its
toric web-diagram.
\begin{figure}[tbph]
\begin{center}
\hspace{-1cm} \includegraphics[width=6cm]{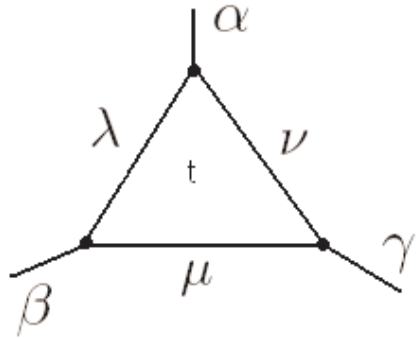}
\end{center}
\par
\vspace{-1cm}
\caption{Toric web-diagram of $\mathcal{O}\left( -3\right) \rightarrow
\mathbb{P}^{2}$ made of three planar vertices}
\label{np2}
\end{figure}
\ \ \ \newline
This local threefold is obtained by gluing \emph{three} 3-vertices. For
simplicity, we consider here also the case where there is no boundaries. The
corresponding partition function reads as follows:
\begin{equation}
Z_{X_{3}}(q,t)=\sum_{\lambda ,\mu ,\nu }C_{\emptyset \mu \nu
^{T}}(q)C_{\lambda \mu ^{T}\emptyset }(q)\text{ }C_{\lambda ^{T}\emptyset
\nu }(q)(-e^{-t})^{\left\vert \nu \right\vert +\left\vert \lambda
\right\vert +\left\vert \nu \right\vert }q^{\kappa (\lambda )+\kappa \left(
\mu \right) +\kappa (\nu )}  \label{np}
\end{equation}%
with
\begin{equation}
\kappa (\lambda )=2\left[ \left( \left\Vert \lambda \right\Vert
^{2}-\left\vert \lambda \right\vert \right) -\left( \left\Vert \lambda
^{T}\right\Vert ^{2}-\left\vert \lambda ^{T}\right\vert \right) \right] ,
\label{k}
\end{equation}%
being the Casimir of the 2d partition.

\section{Beyond the \emph{planar} vertex method}

First, we describe briefly the field theory setting of the local elliptic
curve geometry leaving technical details for next sections. Then we give our
first results regarding the topological non planar 3-vertex formalism and
the explicit expression of the partition function associated with eq(\ref{P1}%
).

\subsection{Field theory set up}

The local CY3 examples we have described above have toric web-diagrams
involving \emph{planar} 3- vertices; see \emph{figures} (\ref{3v})-(\ref{23v}%
)-(\ref{np2}). These toric threefolds $X_{3}$ have a very remarkable field
theory set up; they describe supersymmetric vacua of 2D $\mathcal{N}=2$
linear sigma model with $U^{r}\left( 1\right) $ gauge symmetry and $\left(
r+3\right) $ matter multiplets $\Phi _{i}$,%
\begin{equation}
U^{r}\left( 1\right) :\qquad \Phi _{j}\equiv e^{iq_{j}^{a}}\Phi _{j},\qquad
a=1,...,r,
\end{equation}%
The defining eq of $X_{3}$ is given by the field equation of motion of the $%
D^{a}$ auxiliary fields,%
\begin{equation}
X_{3}:\qquad \frac{\delta \mathcal{L}}{\delta D^{a}}%
=\sum_{i=1}^{r+3}q_{i}^{a}\left\vert z_{i}\right\vert ^{2}=0,\qquad
\label{dt}
\end{equation}%
where the field coordinates $z_{i}$ are the leading components of the chiral
superfields $\Phi _{i}$ and where%
\begin{equation}
\sum_{i=1}^{r+3}q_{i}^{a}=0,\qquad a=1,...,r,
\end{equation}%
stands for the Calabi-Yau condition.

Toric Calabi-Yau threefolds can be also realized as hypersurfaces $H_{3}$ in
higher \emph{d- dimension} complex Kahler toric manifolds $\mathcal{Y}_{d}$,
\begin{equation}
H_{3}\subset \mathcal{Y}_{d},\qquad d\geq 4.  \label{H3}
\end{equation}%
Locally, the Kahler toric d-fold $\mathcal{Y}_{d}$ may be imagined as given
by the toric fibration $R^{d}\times T^{d}$ or as $R^{d}\times F_{d}$ with
fiber $F_{d}=R\times T^{d-1}$. The toric web-diagram of $\mathcal{Y}_{d}$
involve \emph{d}- \emph{dimensional} vertices where shrink all the 1-cycles
of the toroidal fibration.
\begin{figure}[tbph]
\begin{center}
\hspace{-1cm} \includegraphics[width=3cm]{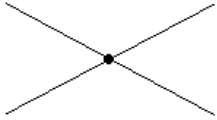}
\end{center}
\par
\vspace{-1cm}
\caption{A generic toric web-diagram of a 4-vertex. In toric geometry, this
web-diagram corresponds to the real base of the local patch $\mathbb{C}^{4}$%
. }
\label{4i}
\end{figure}
\ \ \ \ \newline
The toric CY3 hypersurfaces $H_{3}$ have also a supersymmetric field theory
setting. It will be developed in details in forthcoming sections, see also
the analysis of \textrm{\cite{18}}. The $H_{3}$'s describe as well
supersymmetric vacua of 2D $\mathcal{N}=2$ linear sigma model with gauge $%
\left\{ V_{a}\right\} $ and matter $\left\{ \Phi _{i}\right\} $ multiplets, $%
i=1,...,r+d$. \newline
The defining equation of the toric CY hypersurface $H_{3}$ is given by the
field equations of motion of both the D- and the F- auxiliary fields,%
\begin{equation}
H_{3}:\left\{
\begin{array}{c}
\frac{\delta \mathcal{L}}{\delta D^{a}}=0,\qquad a=1,...,r \\
\frac{\delta \mathcal{L}}{\delta F^{\alpha }}=0,\qquad \alpha =1,...,m%
\end{array}%
\right\vert ,  \label{ft}
\end{equation}%
with $m=d-3$ and $d\geq 4$. The first $r$ equations, which are similar to eq(%
\ref{dt}), reduce the dimension down to $\left( d-r\right) $. The second
equations, which are \emph{gauge invariant} constraint relations,%
\begin{equation}
\left\{
\begin{array}{c}
f_{\alpha }\left( z_{1},...,z_{d}\right) =0 \\
f_{\alpha }\left( \mathrm{\lambda }_{1}z_{1},...,\mathrm{\lambda }%
_{d}z_{d}\right) =0 \\
\mathrm{\lambda }_{j}=e^{iq_{j}^{a}\mathrm{\alpha }_{a}}%
\end{array}%
\right\vert ,\qquad \alpha =1,...,m,
\end{equation}%
reduce the number of free field variables down to 3; say $\left(
w_{1},w_{2},w_{3}\right) $. Up to solving eqs(\ref{ft}), one can express all
the $z_{i}$ field variables in terms of the $w$'s as shown below
\begin{equation}
z_{i}=z_{i}\left( w_{1},w_{2},w_{3}\right) ,\qquad i=1,...,r+d.
\end{equation}%
In the next subsection, we study in details the case $d=4$.

\subsection{Results on\ \emph{non planar} vertex formalism}

The results we will give below concern the following: \newline
(\textbf{1}) \textrm{the toric realization of the local degenerate elliptic
curve }(\ref{P1}), \newline
(\textbf{2}) the set up of the non planar 3- vertex formalism\ and \newline
(\textbf{3}) the computation of the partition function Z$_{H_{3}}$.

\subsubsection{Local degenerate elliptic curve}

Consider the local Calabi-Yau threefold (\ref{P1}) and focus on the
particular local degenerate elliptic curve,
\begin{equation}
H_{3}=\mathcal{O}(+3)\oplus \mathcal{O}(-3)\rightarrow E^{\left( t,\infty
\right) },\qquad m=3.  \label{m}
\end{equation}%
The degenerate elliptic curve $E^{\left( t,\infty \right) }$ is given by the
\emph{toric boundary} (divisor) of the complex projective plane $\mathbb{P}%
^{2}$,
\begin{equation}
E^{\left( t,\infty \right) }=\partial \left( \mathbb{P}^{2}\right) .
\end{equation}%
This is just a compact divisor (hyperline) of $\mathbb{P}^{2}$. The toric
web-diagram associated to (\ref{m}) is given by \emph{figure 5}.\newline
\begin{figure}[tbph]
\begin{center}
\hspace{0cm} \includegraphics[width=4cm]{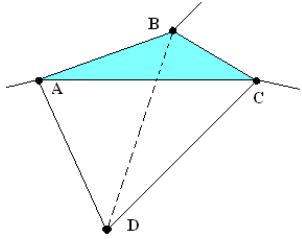}
\end{center}
\par
\vspace{-0.5 cm}
\caption{Non planar toric web-diagram of \ $\mathcal{O}\left( +3\right)
\oplus \mathcal{O}\left( -3\right) \rightarrow E^{\left( t,\infty \right) }$%
. {\protect\small This is a toric CY3 divisor of the four dimension complex
Kahler manifold} $\mathcal{O}\left( +3\right) \oplus \mathcal{O}\left(
-3\right) \rightarrow P^{2}$ \ {\protect\small The hollow triangle ABC
refers to the degenerate elliptic curve }$E^{\left( t,\infty \right) }$%
{\protect\small . The full triangles ABD, ZCD, BCD refer to the three other
projective planes. }}
\label{4v2}
\end{figure}
\ \ \newline
The non compact toric 4- fold $\mathcal{Y}_{4}$ of eq(\ref{H3}) is given by%
\begin{equation}
\mathcal{Y}_{4}=\mathcal{O}(-3)\rightarrow WP_{1,1,1,3}^{3},
\end{equation}%
where $WP_{1,1,1,3}^{3}$ stands for the complex 3- dimension weighted
projective space. To keep in touch with the Calabi-Yau condition, we promote
$\mathcal{Y}_{4}$ to the toric Calabi-Yau 4-fold%
\begin{equation}
X_{4}=\mathcal{O}(-6)\rightarrow WP_{1,1,1,3}^{3},
\end{equation}%
and in general to
\begin{equation}
X_{4}=\mathcal{O}(-3-m)\rightarrow WP_{1,1,1,m}^{3}
\end{equation}%
with $m\geq 1$.

\subsubsection{Toric cap and toric cylinder}

From eq(\ref{m}), one distinguishes two special divisors of the local
degenerate elliptic curve $H_{3}$:\newline
(1) \emph{"toric cap": see figure 6}\newline
This divisor corresponds to $H_{3}$ taken as the fibration $\mathcal{O}%
(-3)\rightarrow Y_{2}$. \ The base $Y_{2}$ is a compact complex surface
given by
\begin{equation}
Y_{2}=\mathcal{O}(+3)\rightarrow E^{\left( t,\infty \right) }.  \label{tc}
\end{equation}%
The toric web-diagram of the complex surface $Y_{2}$ is exhibited in\emph{\
figure 5:}\newline

\begin{figure}[tbph]
\begin{center}
\hspace{0cm} \includegraphics[width=5cm]{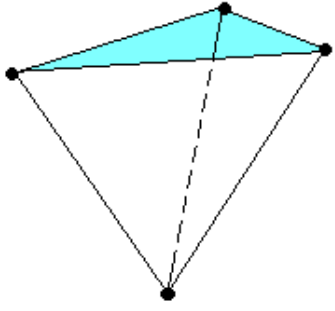}
\end{center}
\par
\vspace{-1cm}
\caption{{\protect\small Toric web-diagram of} \ $\mathcal{O}\left(
+3\right) \rightarrow E^{\left( t,\infty \right) }$. {\protect\small This}
{\protect\small figure looks like a "toric cap" obtained by gluing three
triangles as shown on the figure.}}
\end{figure}
\ \ \newline
The first Chern class $c_{1}\left( T^{\ast }Y_{2}\right) $ is equal to $+3$.
The toric web-diagram of $E^{\left( t,\infty \right) }$ is given by the
boundary of the triangle and $\mathcal{O}(+3)$ is a compact line. \newline
The toric web-diagram of $Y_{2}$ is non planar and can be thought of as the
triangulation of the topological cap of \textrm{\cite{9}}. Below, we will
refer to $Y_{2}$ as the \emph{"toric cap"}. Notice also the following
features:\newline
(\textbf{a}) the complex surface $Y_{2}$ is a compact divisor of $H_{3}$; it
is made of the union of three complex projective planes, which we denote as $%
\mathbb{P}_{1}^{2}$, $\mathbb{P}_{2}^{2}$ and $\mathbb{P}_{3}^{2}$. \newline
The projective planes $\mathbb{P}_{i}^{2}$ belong to three different $%
\mathbb{C}^{3}$ spaces of the ambient $\mathbb{C}^{4}$.\newline
(\textbf{b}) the toric web-diagram of $Y_{2}$ is made of three triangles as
shown\textrm{\footnote{$Y_{2}$ can be imagined as the triangulation of the
cap.}} in the \emph{figure 6}. Recall that the toric web-diagram of a
generic projective plane is given by a triangle (\emph{figure 3}).\newline
The projective planes (triangles) have mutual intersections $I_{ij}$ along
complex projective lines (edges) and a \emph{non planar} tri-intersection
vertex $I_{123}$.\newline
(2) \emph{toric cylinder}\newline
This divisor corresponds to think about $H_{3}$ as the fibration $\mathcal{O}%
(+3)\rightarrow \widetilde{Y}_{2}$.\newline
Here the base $\widetilde{Y}_{2}$ is a non compact complex surface given by
\begin{equation}
\widetilde{Y}_{2}=\mathcal{O}(-3)\rightarrow E^{\left( t,\infty \right) }.
\label{ct}
\end{equation}%
Its first Chern class $c_{1}\left( T^{\ast }\widetilde{Y}_{2}\right) $ is
equal to $-3$. Here also the toric web-diagram of $E^{\left( t,\infty
\right) }$ is the boundary of the triangle and $\mathcal{O}(-3)$ is a non
compact line. The toric web-diagram of $\widetilde{Y}_{2}$ is \emph{non
planar}; it can be thought of as the triangulation of the cylinder $\mathbb{R%
}\times \mathbb{S}^{1}$. We will refer to $\widetilde{Y}_{2}$ as the \emph{%
"toric cylinder"} whose toric web-diagram is shown in the \emph{figure }\ref%
{cyl3}.\newline
\begin{figure}[tbph]
\begin{center}
\hspace{-1cm} \includegraphics[width=10cm]{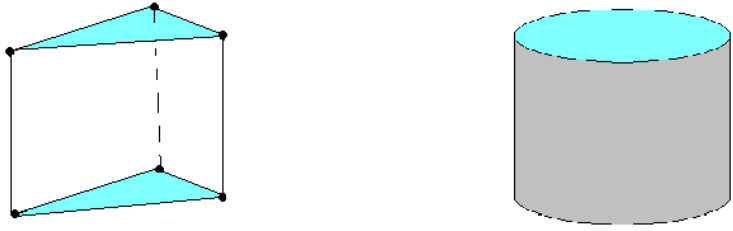}
\end{center}
\par
\vspace{-1cm}
\caption{{\protect\small On left, the real skeleton of the toric web-diagram
of the fibration of }$O\left( -3\right) \rightarrow E^{\left( t,\infty
\right) }${\protect\small . It could be interpreted as a triangulation of
the cylinder of base }$E^{\left( t,\mathrm{\protect\mu }\right) }$%
{\protect\small , given by the boundary of a triangle, and a non compact
line as a fiber. On right, the usual cylinder }$R\times S^{1}$%
{\protect\small .}}
\label{cyl3}
\end{figure}
\ \ \ \newline
The $\widetilde{Y}_{2}$ divisor of $H_{3}$ is made of the union of three
sheets \emph{O}$\left( -3\right) \rightarrow \mathbb{P}_{i}^{1}$ belonging
to three different $\mathbb{C}^{3}$ spaces of the ambient $\mathbb{C}^{4}$.%
\newline
From eqs(\ref{tc}) and (\ref{ct}), it is clear that the elliptic curve $%
E^{\left( t,\infty \right) }$ is an intersecting curve of the complex
surfaces $Y_{2}$ and $\widetilde{Y}_{2}$.

\subsection{Topological partition function}

To build the \emph{non planar} 3-vertex formalism for the local elliptic
curve $H_{3}$, we will follow the construction used in the derivation of the
usual topological 3-vertex method \cite{9}. \newline
However since the local elliptic curve is a CY3 hypersurface in $X_{4}$,
\begin{equation}
H_{3}\subset X_{4},
\end{equation}%
with%
\begin{equation}
X_{4}=\mathcal{O}(-3-m)\rightarrow WP_{1,1,1,m}^{3},
\end{equation}%
a convenient way to achieve the goal is to proceed as follows:\newline
(\textbf{1}) develop the 4-vertex formalism for the ambient toric CY4-fold $%
X_{4}$.\newline
(\textbf{2}) compute the partition function for $X_{4}$. Actually this step
may be also viewed as alternative way to get the 4d generalization of the
MacMahon function\textrm{\ \cite{DJS}}. \newline
(\textbf{3}) impose the appropriate constraint relations to get the \emph{%
non planar }3-vertex and the topological partition function for the local
elliptic curve H$_{3}$.\newline
It is interesting to note here the emergence of a 4-vertex formalism in the
construction. This is not surprising since the toric web-diagrams of $H_{3}$
and $X_{4}$ have quite similar skeletons. In the first case the tetrahedron
is hollow and in the second it is full.

The first step to realize these objectives is specify the special lagrangian
fibration of the toric CY4-fold like $X_{4}\sim R^{4}\times F_{4}$ with
fiber taken as%
\begin{equation}
F_{4}\sim R\times T^{3}.
\end{equation}%
On the hypersurface $H_{3}$ in $X_{4}$, real 1-cycles of $F_{4}$ shrink and
one is left with the usual special lagrangian fibration of the toric
CY3-folds $H_{3}\sim R^{3}\times F_{3}$ with
\begin{equation}
F_{3}\sim R\times T^{2}.
\end{equation}%
In \cite{20}, we give the explicit expressions of the various hamiltonians
and the values of the vertices of the web-diagrams solving the Calabi-Yau
conditions.\newline
The next step is to study the 4-vertex formalism of $X_{4}$ and its
reduction down to the toric hypersurface $H_{3}$.

\subsubsection{Toric web-diagrams and generalized partitions}

The toric web-diagrams of $X_{4}$ and $H_{3}$ have been described above (%
\emph{figure 5}). For the case $X_{4}$, the toric web-diagram can be
decomposed into \emph{four} local patches
\begin{equation}
\mathcal{U}_{1},\quad \mathcal{U}_{2},\quad \mathcal{U}_{3},\quad \mathcal{U}%
_{4}.
\end{equation}%
To each patch $\mathcal{U}_{i}\sim \mathbb{C}^{4}$ with fibration $%
R^{4}\times F_{4}$ it is associated a topological 4-vertex $\mathcal{C}%
_{\left( 4\right) }$. This vertex depends on the boundary conditions on its
external legs. We will see that, using 3d generalized Young diagrams, it can
be either defined as
\begin{equation}
\mathcal{C}_{\left( 4\right) }=C_{\mathrm{\Lambda \Sigma \Upsilon \Gamma }},
\end{equation}%
or equivalently by using 2d partition like
\begin{equation}
\mathcal{C}_{\left( 4\right) }=\mathcal{C}_{\left( \mathrm{\alpha \beta
\gamma }\right) \left( \mathrm{\delta \epsilon \varepsilon }\right) \left(
\mathrm{\zeta \eta \theta }\right) \left( \mathrm{\lambda \mu \nu }\right) }.
\label{vcc}
\end{equation}%
The toric web-diagram of $H_{3}$ is induced from the one of $X_{4}$. It can
be also decomposed into \emph{four} local patches as follows,%
\begin{equation}
\mathcal{U}_{1}^{\ast },\quad \mathcal{U}_{2}^{\ast },\quad \mathcal{U}%
_{3}^{\ast },\quad \mathcal{U}_{4}^{\ast },
\end{equation}%
where the \emph{asterix} refers to the projection
\begin{equation}
\ast :X_{4}\rightarrow H_{3},\qquad \mathcal{U}_{i}\rightarrow \mathcal{U}%
_{i}^{\ast }.
\end{equation}%
To each patch $\mathcal{U}_{i}^{\ast }\sim \mathbb{C}^{3}$ with fibration $%
R^{3}\times F_{3}$ it is associated a \emph{non planar} topological 3-vertex
$\mathcal{C}_{\left( 3\right) }^{\ast },$%
\begin{equation}
\ast :\mathcal{C}_{\left( 4\right) }\rightarrow \mathcal{C}_{\left( 3\right)
}^{\ast }.  \label{st}
\end{equation}%
To get the 4- vertex $C_{\left( 4\right) }$, the partition function $%
Z_{X_{4}}$ and the topological partition function $Z_{H_{3}}$ of the local
elliptic curve, we need first introducing some key tools. \newline
In the standard 3-vertex formalism of \cite{9,11}\textrm{,} one uses a set
of basic objects; in particular 2d- and 3d- partitions. In the 4-vertex
formalism, we have to build the analogue of these mathematical ingredients.

\paragraph{$\mathbf{\protect\alpha }$\emph{) 3d partitions }\newline
}

Roughly, a 3d partition $\Pi $ can be thought of as an integral $N_{1}\times
N_{2}$ rank two tensor $\left( \Pi _{ia}\right) $ with the property,
\begin{equation}
\Pi =\left\{ \Pi _{i,a}\in \mathbb{Z}_{+},\qquad \Pi _{i,a}\geq \Pi
_{i+j,a+b}\geq 0\right\} ,  \label{3p}
\end{equation}%
where $i,$ $j=1,2,...,N_{1}$ and $a,$ $b=1,2,...,N_{2}$. \newline
The 3d partition, which has been used for various purposes, has a set of
remarkable combinatorial features. Below, we give useful ones. \newline
(\textbf{i}) 3d partitions are generalizations of the usual 2d partitions $%
\lambda =\left( \lambda _{1},\lambda _{2},...\right) $ with $\lambda
_{1}\geq \lambda _{2}\geq ....\geq 0$ and the integers $\lambda _{i}\in
\mathbb{Z}_{+}$.\newline
By setting $N_{3}=\Pi _{1,1}$, the 3d partitions can be imagined as a cubic
sublattice of $\mathbb{Z}_{+}^{3}$
\begin{equation}
\left[ 1,N_{1}\right] \times \left[ 1,N_{2}\right] \times \left[ 1,N_{3}%
\right] .
\end{equation}%
The cubic diagram of $\Pi $ can be considered as a set of unit cubes $\left(
i,j,k\right) $ with integer coordinates such that $\left( i,j\right) \in
\lambda $ and $1\leq k\leq \Pi (i,j)$. The integers $\Pi (i,j)$ define the
height of the stack of cubes on the $\left( x_{1},x_{2}\right) $ plane. The
projection of $\Pi $ on the $\left( x_{1},x_{2}\right) $ plane is just the
2d partition $\lambda $. \newline
(\textbf{ii}) A subclass of 3d partitions solving the conditions (\ref{3p})
is given by the particular representation%
\begin{equation}
\Pi =\lambda \otimes \mu ,\qquad \Pi _{ia}=\lambda _{i}\mu _{a},\qquad
\lambda _{i}\mu _{a}\geq \lambda _{i+j}\mu _{a+b},
\end{equation}%
where $\lambda $ and $\mu $ are 2d partitions as in eq(\ref{2p}).\newline
We also have the following associated ones:
\begin{equation}
\Pi ^{\text{{\small T}}}=\lambda ^{\text{{\small T}}}\otimes \mu ,\qquad
\widetilde{\Pi }=\lambda \otimes \mu ^{\text{{\small T}}},\qquad \widetilde{%
\Pi ^{\text{{\small T}}}}=\lambda ^{\text{{\small T}}}\otimes \mu ^{\text{%
{\small T}}},
\end{equation}%
where $\lambda ^{\text{{\small T}}}$ stands for the usual transpose of the
Young diagram $\lambda $.\newline
(\textbf{iii}) Like in the case of 2d partitions, one may associate to each
3d partition $\Pi $ a Fock space state $\left\vert \Pi _{ia}\right\rangle $
with norm $\left\langle \Pi |\Pi \right\rangle \equiv \left\Vert \Pi
\right\Vert ^{2}$,%
\begin{equation}
\left\Vert \Pi \right\Vert ^{2}=\sum_{i\geq 1}\left( \sum_{a\geq 1}\left(
\Pi _{ia}\right) ^{2}\right) =\sum_{a\geq 1}\left( \sum_{i\geq 1}\left( \Pi
_{ia}\right) ^{2}\right) .  \label{tra}
\end{equation}%
$\left\langle \Pi _{ia}\right\vert $ stands for the dual state associated
with the dual partition $\Pi ^{\mathrm{+}}=\widetilde{\Pi ^{\text{{\small T}}%
}}$. We also have the following relation
\begin{equation}
I_{id}=\sum_{3d\text{ partitions}}\left\vert \Pi _{ia}\right\rangle
\left\langle \Pi _{ia}\right\vert ,\qquad \left\langle \Pi _{ia}|\Pi
_{jb}\right\rangle =\delta _{ij}\delta _{ab},
\end{equation}%
defining the resolution of the identity operator $I_{id}$.\newline
(\textbf{iv}) The number $\left\vert \Pi \right\vert $ of unit boxes (cubes)
of the 3d partition is defined as
\begin{equation}
\left\vert \Pi \right\vert =\sum_{i,a}\Pi _{i,a}.
\end{equation}%
(\textbf{v}) The boundary $\left( \partial \Pi \right) $ of the 3d partition
$\Pi $ is given by the 2d profile of the corresponding generalized Young
diagram. As this property is important for the present study, let give some
details.\newline
Given a 3d partition $\Pi $, the \emph{boundary term }on\emph{\ }the plane $%
x_{i}=N_{i}$ is a Young diagram (2d partition). On the planes $x_{1}=N_{1}$,
$x_{2}=N_{2}$ and $x_{3}=N_{3}$, the boundary of $\Pi $ is composed of by
three 2d partitions $\lambda ,$ $\mu $ and $\nu $. So we then have:%
\begin{equation}
\partial \Pi =\left( \lambda ,\mu ,\nu \right) .
\end{equation}%
Particular boundaries are given by the case where a 2d partition is located
at infinity; that is there is no boundary. We distinguish the following
situations:%
\begin{eqnarray}
\partial \Pi &=&\left( \emptyset ,\mu ,\nu \right) ,  \notag \\
\partial \Pi &=&\left( \emptyset ,\emptyset ,\nu \right) , \\
\partial \Pi &=&\left( \emptyset ,\emptyset ,\emptyset \right) ,  \notag
\end{eqnarray}%
where $\emptyset $ stands for the vacuum.\newline
(\textbf{vi}) A convenient way to deal with 3d partitions is to slice them
as a sequence of 2d partitions with interlacing relations. We mainly
distinguish two kinds of sequences of 2d partitions: perpendicular and
diagonal. We will not need this property here; but for details on this
matter see for instance \textrm{\cite{11}} and refs therein.

\paragraph{$\mathbf{\protect\beta }$\emph{)} \emph{4d partitions }\newline
}

The 4d partitions $\mathcal{P}$ are extensions of the 3d partitions $\Pi $
considered above. They can be imagined as \emph{4d generalized} Young
diagrams described by the typical integral rank 3- tensor ,%
\begin{equation}
\mathcal{P}_{ia\alpha }\in \mathbb{Z}_{+},\qquad \text{with\qquad }\mathcal{P%
}_{ia\alpha }\geq \mathcal{P}_{\left( i+j\right) \left( a+b\right) \left(
\alpha +\beta \right) },\quad
\end{equation}%
with $1\leq i\leq N_{1},$ $1\leq a\leq N_{2}$ and $1\leq \alpha \leq N_{3}$.
\newline
Several properties of 2d and 3d partitions extend to the 4d case; there are
also specific properties in particular those concerning their slicing into
lower dimensional ones. Below we describe some particular properties of 4d
partitions by considering special representations. \newline
Sub-classes of 4d partitions are given by:\newline
(\textbf{i}) the product of a 2d- and a 3d- partitions $\mu $ and $\Pi $\
like%
\begin{equation}
\mathcal{P}=\mu \otimes \Pi ,\qquad \left( \mathcal{P}_{ia\alpha }\right)
=\left( \mu _{i}\Pi _{a\alpha }\right) ,
\end{equation}%
with $i=1,...,N_{1}$; $a=1,...,N_{2}$ and $\alpha =1,...,N_{3}$.\newline
(\textbf{ii}) the product of three kinds of 2d- partitions.%
\begin{equation}
\mathcal{P}=\lambda \otimes \mu \otimes \nu ,\qquad \left( \mathcal{P}%
_{ia\alpha }\right) =\left( \lambda _{i}\mu _{a}\nu _{\alpha }\right) .
\label{mn}
\end{equation}%
The boundary $\partial \mathcal{P}$ of a generic 4d partition $\mathcal{P}$
can be defined in two ways. First in terms of 3d partitions as follows%
\begin{equation}
\partial \mathcal{P}=\left( \Lambda ,\Pi ,\Sigma ,\Upsilon \right) .
\label{3d}
\end{equation}%
We also have the following particular boundary conditions%
\begin{eqnarray}
\text{case I} &:&\partial \mathcal{P}=\left( \varnothing ,\Pi ,\Sigma
,\Upsilon \right) ,  \notag \\
\text{case II} &:&\partial \mathcal{P}=\left( \varnothing ,\varnothing
,\Sigma ,\Upsilon \right) ,  \notag \\
\text{case III} &:&\partial \mathcal{P}=\left( \varnothing ,\varnothing
,\varnothing ,\Upsilon \right) ,  \label{cas} \\
\text{case IV} &:&\partial \mathcal{P}=\left( \varnothing ,\varnothing
,\varnothing ,\varnothing \right) ,  \notag
\end{eqnarray}%
where $\varnothing $ stands for the the \emph{"3d vacuum"} (no boundary
condition).\newline
Second by using 2d partitions to define boundary of $\mathcal{P}$ like%
\begin{equation}
\partial \mathcal{P}=\left( \text{ }\left[ \mathrm{a,b,c}\right] ;\text{ \ }%
\left[ \mathrm{d,e,f}\right] ;\text{ }\left[ \mathrm{g,h,i}\right] ;\text{ }%
\left[ \mathrm{j,k,l}\right] \text{ \ }\right) ,  \label{dpp}
\end{equation}%
where $\left[ \mathrm{abc}\right] $, ...and $\left[ \mathrm{jkl}\right] $
stand for the boundaries of the 3d partitions $\Lambda ,...$ and $\Upsilon $.%
\newline
Notice that the second representation is more richer since along with the
configuration%
\begin{equation}
\varnothing =\left[ \emptyset ,\emptyset ,\emptyset \right] ,
\end{equation}%
we have moreover the two following extra configurations%
\begin{equation}
\left[ \mathrm{\emptyset ,b,c}\right] ,\qquad \left[ \mathrm{\emptyset
,\emptyset ,c}\right] .
\end{equation}%
For the case I of eq(\ref{cas}) corresponds then the three following
boundary configurations%
\begin{equation}
\partial \mathcal{P}=\left\{
\begin{array}{c}
\text{case i}:\qquad \left( \left[ \mathrm{\emptyset ,b,c}\right] ;\text{%
\quad }\left[ \mathrm{d,e,f}\right] ;\text{\quad }\left[ \mathrm{g,h,i}%
\right] ;\text{\quad }\left[ \mathrm{j,k,l}\right] \right) \\
\\
\text{case ii}:\qquad \left( \left[ \mathrm{\emptyset ,\emptyset ,c}\right] ;%
\text{\quad }\left[ \mathrm{d,e,f}\right] ;\text{\quad }\left[ \mathrm{g,h,i}%
\right] ;\text{\quad }\left[ \mathrm{j,k,l}\right] \right) \\
\\
\text{case iii}:\qquad \left( \left[ \mathrm{\emptyset ,\emptyset ,\emptyset
}\right] ;\text{\quad }\left[ \mathrm{d,e,f}\right] ;\text{\quad }\left[
\mathrm{g,h,i}\right] ;\text{\quad }\left[ \mathrm{j,k,l}\right] \right)%
\end{array}%
\right\vert ,
\end{equation}%
where the last one (case iii) is the case I described by the first relation
of eq(\ref{cas}). \newline
This property indicates that one disposes of different ways to deal with 4d
partitions either the simplest one using 3d-partitions or the more refined
on involving 2d partitions. Below we consider both representations.

Notice moreover that given a 4d partition $\mathcal{P}$, we can associate to
it various kinds of transpose partitions. Using the particular realization
eq(\ref{mn}), the corresponding transposes read as
\begin{eqnarray}
&&\lambda ^{\mathrm{T}}\otimes \mu \otimes \nu ,\qquad \lambda \otimes \mu ^{%
\mathrm{T}}\otimes \nu ,\qquad \lambda \otimes \mu \otimes \nu ^{\mathrm{T}},
\notag \\
&&\lambda ^{\mathrm{T}}\otimes \mu ^{\mathrm{T}}\otimes \nu ,\qquad \lambda
\otimes \mu ^{\mathrm{T}}\otimes \nu ^{\mathrm{T}},\qquad \lambda ^{\mathrm{T%
}}\otimes \mu \otimes \nu ^{\mathrm{T}}, \\
&&\lambda ^{\mathrm{T}}\otimes \mu ^{\mathrm{T}}\otimes \nu ^{\mathrm{T}},
\notag
\end{eqnarray}%
where $\lambda ^{\mathrm{T}}$ stands for the usual transpose of the Young
diagram $\lambda $. \newline
The exact mathematical definitions and the full properties of 4d partitions
are not our immediate objective here; they need by themselves a separate
study. Here above we have given just the needed properties to set up the
structure of the 4-vertex formalism and its restricted \emph{non planar}
3-vertex method.

\subsubsection{Tetra- vertex $C_{\left( 4\right) }$ and $Z_{X_{4}}$}

\paragraph{$\mathbf{\protect\alpha }$\emph{)} \emph{the} \emph{4- vertex }$%
C_{\left( 4\right) }$ \newline
}

The 4- vertex $C_{\left( 4\right) }$\ of the toric Calabi-Yau X$_{4}$ can be
built by extending the 3-vertex construction eq(\ref{c3}).

In the 3d partition set up, the vertex $\mathcal{C}_{\left( 4\right) }$ has
four external legs $L_{\Lambda },L_{\Sigma },L_{\Upsilon },L_{\Gamma }$ with
boundary conditions as in eq(\ref{3d}). The 4-vertex $\mathcal{C}_{\left(
4\right) }$ with boundary conditions $\left( \Lambda \Sigma \Upsilon \Gamma
\right) $ can be defined as a function of the Bolzmann weight $q=e^{-\beta }$
as follows:
\begin{equation}
\mathcal{C}_{\Lambda \Sigma \Upsilon \Gamma }\equiv \mathcal{C}_{\Lambda
\Sigma \Upsilon \Gamma }\left( q\right) .
\end{equation}%
In the case where $\Lambda =\Sigma =\Upsilon =\Gamma =\varnothing $, the
corresponding 4- vertex $\mathcal{C}_{\varnothing \varnothing \varnothing
\varnothing }$ should be equal to the generating function $Z_{\mathbb{C}%
^{4}} $ of the 4d generalized Young diagrams
\begin{equation}
\mathcal{C}_{\varnothing \varnothing \varnothing \varnothing }=Z_{\mathbb{C}%
^{4}}.
\end{equation}%
Recall that the generating functional $Z_{\mathbb{C}^{4}}$ can be defined as
a power series like,
\begin{equation}
Z_{\mathbb{C}^{4}}=\sum_{\text{4d partitions }\mathcal{P}}q^{\left\vert
\mathcal{P}\right\vert },
\end{equation}%
where $\left\vert \mathcal{P}\right\vert $ is the number of hypercubes in $%
\mathcal{P}$.\newline
In the generic case, $\mathcal{C}_{\Lambda \Sigma \Upsilon \Gamma }$ should
be given by the generalization\footnote{%
the partition function $Z_{C^{3}}$ can be also defined as the generating
function of $3d$ partitions \cite{11}.} of the 3-vertex (\ref{c}) and could
a priori be expressed in terms of the product of some hypothetic generalized
Schur functions $\mathcal{S}_{\Lambda }\left( q\right) $.

In the 2d partition set up, one can $\mathcal{C}_{\left( 4\right) }$ in
quite similar manner. Using eqs(\ref{3d}-\ref{dpp}), we can generally define
it as in eq(\ref{vcc}). This is a kind of rank 12 object
\begin{equation}
\mathcal{C}_{\left( \mathrm{abc}\right) \left( \mathrm{def}\right) \left(
\mathrm{ghi}\right) \left( \mathrm{jkl}\right) },
\end{equation}%
depending on the Boltzmann weight and the boundary conditions (external
momenta) $\mathrm{a,...,l}$. \newline
To obtain its explicit expression, we use the following\newline
(\textbf{i}) the relation between the 4-vertex and composites of \emph{planar%
} 3-vertices.\newline
(\textbf{ii}) the results on the usual 3-vertex formalism. \newline
The first property follows by noting that 4-vertices of the toric
web-diagram of $X_{4}$ with the special Lagrangian fibration
\begin{equation}
R^{4}\times R\times T^{3},
\end{equation}%
corresponds to the intersection of the \emph{planar 3-vertices} of three
triangles. To fix the ideas, consider \emph{figure 5} and focus on the point
A representing a 4-vertex of the toric web-diagram of X$_{4}$. The point A
is the intersection%
\begin{equation}
A=\Delta _{1}\cap \Delta _{2}\cap \Delta _{3},  \label{a}
\end{equation}%
of the triangles,%
\begin{eqnarray}
\Delta _{1} &=&\text{ triangle ABC,}  \notag \\
\Delta _{2} &=&\text{ triangle ABD,}  \label{d} \\
\Delta _{3} &=&\text{ triangle ACD,}  \notag
\end{eqnarray}%
These triangles are boundary faces of the tetrahedron
\begin{equation}
\text{ABCD}.
\end{equation}%
Inside of the tetrahedron, the toric fiber is
\begin{equation}
T^{3}=\mathbb{S}^{1}\times \mathbb{S}^{1}\times \mathbb{S}^{1}.
\end{equation}%
On each triangle face, a circle shrinks leaving $T^{2}$. \newline
On each egde of a triangle, one more circle shrinks leaving $\mathbb{S}^{1}$.%
\newline
At the vertex A, all 1-cycles of $T^{3}$ shrinks down to zero.\newline
The property captured by eqs(\ref{a}) means that we may relate the 4-vertex $%
C_{\Lambda \Sigma \Upsilon \Gamma }$ to three \emph{planar }vertices of the
triangles (\ref{d}). This can be done by expressing the 3d partitions $%
\left( \Lambda ,\Sigma ,\Upsilon ,\Gamma \right) $ in terms of 2d partitions
$\left( \mathrm{a,b,c}\right) ,$ $\left( \mathrm{d,e,f}\right) ,$ $\left(
\mathrm{g,h,i}\right) $ and $\left( \mathrm{j,k,l}\right) $ as follows%
\begin{eqnarray}
\Lambda &=&\left( \mathrm{a,d,g}\right) ,\qquad \Sigma =\left( \mathrm{%
b,c,\emptyset }\right) ,  \notag \\
\Upsilon &=&\left( \mathrm{e,\emptyset ,f}\right) ,\qquad \Gamma =\left(
\mathrm{\emptyset ,h,i}\right) .  \label{lam}
\end{eqnarray}%
The decomposition (\ref{lam}) is illustrated on the \emph{formal} figure \ref%
{066} where the three triangles are represented in different colors. \newline
\begin{figure}[h]
\begin{center}
\hspace{-1cm} \includegraphics[width=6cm]{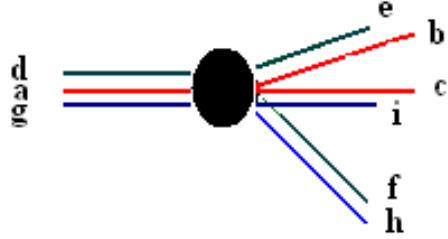}
\end{center}
\par
\vspace{-1cm}
\caption{{\protect\small A typical 4-vertex in }$\mathcal{O}\left(
-3-m\right) \rightarrow WP_{1,1,1,m}^{3}$ using 2d partitions. This is a
spatial vertex made of three planar 3-vertices: ($\mathrm{abc),(def)}$ and ($%
\mathrm{ghi}$). }
\label{066}
\end{figure}
\ \newline
Substituting $\left( \Lambda \Sigma \Upsilon \Gamma \right) $ as in eq(\ref%
{lam}), we can first rewrite $\mathcal{C}_{\Lambda \Sigma \Upsilon \Gamma }$
like $\mathcal{C}_{\left( \mathrm{adh}\right) \left( \mathrm{bc\emptyset }%
\right) \left( \mathrm{e\emptyset f}\right) \left( \mathrm{\emptyset hi}%
\right) }$. The latter reads immediately from the \emph{figure 8} and is
given by
\begin{equation}
\mathcal{C}_{\left( \mathrm{adg}\right) \left( \mathrm{bc\emptyset }\right)
\left( \mathrm{e\emptyset f}\right) \left( \mathrm{\emptyset hi}\right) }=C_{%
\mathrm{abc}}C_{\mathrm{def}}C_{\mathrm{ghi}},  \label{cabc}
\end{equation}%
where $C_{\mathrm{abc}}$, $C_{\mathrm{def}}$ and $C_{\mathrm{ghi}}$ are
topological 3-vertices with the explicit expression \textrm{eq(\ref{c})}.

\paragraph{$\mathbf{\protect\beta }$\emph{)} \emph{the function} $Z_{X_{4}}$%
\newline
}

\begin{figure}[h]
\begin{center}
\hspace{-1cm} \includegraphics[width=6cm]{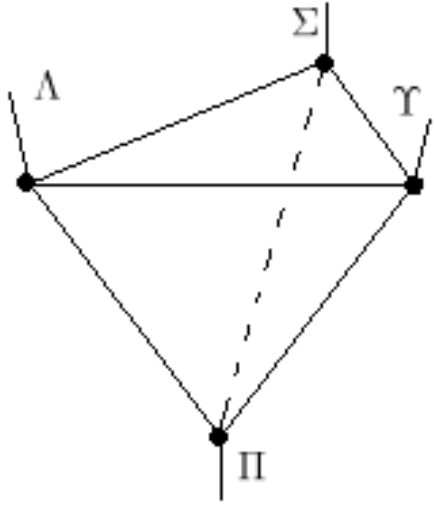}
\end{center}
\par
\vspace{-1cm}
\caption{A typical {\protect\small the toric web-diagram of X}$_{4}$%
{\protect\small \ with boundary conditions }$\left( \Lambda \Pi \Sigma
\Upsilon \right) .$}
\label{06}
\end{figure}

The the toric web-diagram of the $X_{4}$ 4-fold is given by \emph{figure}
\emph{\ref{06}}.\newline
The corresponding partition function $Z_{X_{4}}$ can be computed by
specifying the 4-vertices, propagators, framings and using Feynman like
rules. Notice that in the 2d partition set up, the toric webs of $X_{4}$ and
$H_{3}$ are as in the \emph{figure 10.}

\begin{figure}[tbph]
\begin{center}
\hspace{-1cm} \includegraphics[width=12cm]{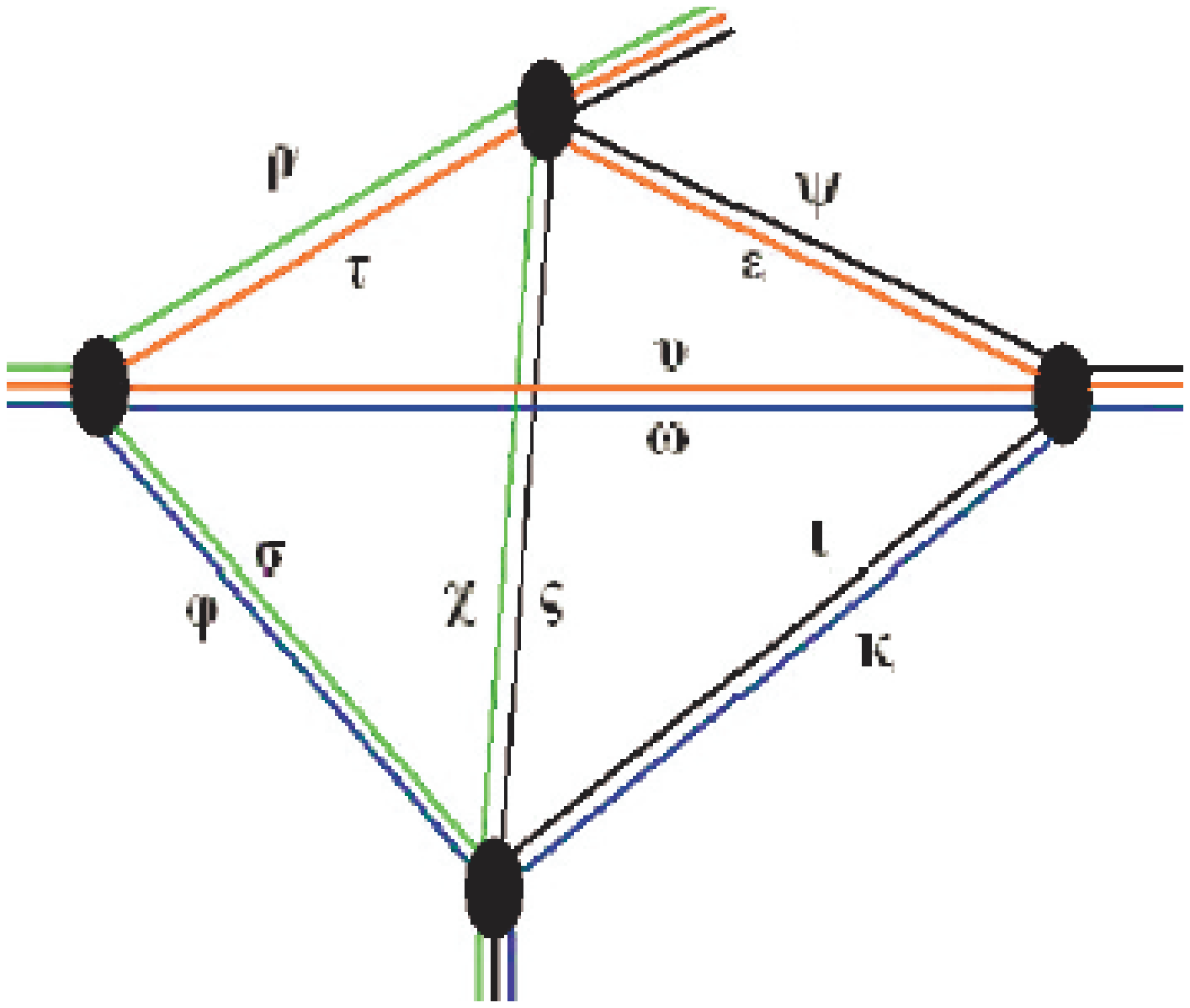}
\end{center}
\par
\vspace{-1cm}
\caption{{\protect\small toric web-diagram of the local elliptic curve using
2d partitions. External and }internal momenta have been expressed in terms
of 2d partitions.{\protect\small \ }}
\end{figure}
\ \newline
Using the momenta prescriptions described by the Young diagrams of the \emph{%
figure 10 }as\emph{\ }well as\emph{\ }trivial boundary conditions for the
extrenal legs, the partition function reads in terms of the Kahler modulus $%
t $ of $X_{4}$ as follows:%
\begin{equation}
Z_{H_{3}}=\sum_{\left\{ \varkappa \right\} }\left[ \left( \mathcal{A}_{\tau
\nu ^{\mathrm{T}}\omega \varphi ^{\mathrm{T}}\rho \sigma ^{\mathrm{T}%
}}\right) \left( \mathcal{B}_{\varepsilon \tau ^{\mathrm{T}}\chi \rho ^{%
\mathrm{T}}\psi \varsigma ^{\mathrm{T}}}\right) \left( \mathcal{F}_{\upsilon
\epsilon ^{\mathrm{T}}\iota \psi ^{\mathrm{T}}\kappa \omega ^{\mathrm{T}%
}}\right) \left( \mathcal{G}_{\varphi \kappa ^{\mathrm{T}}\chi \sigma ^{%
\mathrm{T}}\iota \varsigma ^{\mathrm{T}}}\right) \mathcal{H}_{\tau \upsilon
\omega \varphi \rho \sigma \varepsilon \chi \iota \kappa \psi \varsigma }%
\right]
\end{equation}%
where the sum over $\left\{ \varkappa \right\} $ stands for the collective
sum over the 2d- partitions $\varkappa =\tau ,$ $\upsilon ,$ $\omega ,$ $%
\varphi ,$ $\rho ,$ $\sigma ,$ $\varepsilon ,$ $\chi ,$ $\iota ,$ $\kappa ,$
$\psi ,$ $\varsigma $ and where we have set
\begin{eqnarray}
\mathcal{A}_{\tau \upsilon ^{\mathrm{T}}\omega \varphi ^{\mathrm{T}}\rho
\sigma ^{\mathrm{T}}} &=&C_{\emptyset \tau \upsilon ^{\mathrm{T}%
}}C_{\emptyset \omega \varphi ^{\mathrm{T}}}C_{\emptyset \rho \sigma ^{%
\mathrm{T}}},  \notag \\
\mathcal{B}_{\varepsilon \tau ^{\mathrm{T}}\chi \rho ^{\mathrm{T}}\psi
\varsigma ^{\mathrm{T}}} &=&C_{\emptyset \varepsilon \tau ^{\mathrm{T}%
}}C_{\emptyset \chi \rho ^{\mathrm{T}}}C_{\emptyset \psi \varsigma ^{\mathrm{%
T}}},  \notag \\
\mathcal{F}_{\upsilon \varepsilon ^{\mathrm{T}}\iota \psi ^{\mathrm{T}%
}\kappa \omega ^{\mathrm{T}}} &=&C_{\emptyset \upsilon \varepsilon ^{\mathrm{%
T}}}C_{\emptyset \iota \psi ^{\mathrm{T}}}C_{\emptyset \kappa \omega ^{%
\mathrm{T}}}, \\
\mathcal{G}_{\varphi \kappa ^{\mathrm{T}}\chi \sigma ^{\mathrm{T}}\iota
\varsigma ^{\mathrm{T}}} &=&C_{\emptyset \varphi \kappa ^{\mathrm{T}%
}}C_{\emptyset \chi \sigma ^{\mathrm{T}}}C_{\emptyset \iota \varsigma ^{%
\mathrm{T}}},  \notag
\end{eqnarray}%
and
\begin{equation}
\mathcal{H}_{\tau \upsilon \omega \varphi \rho \sigma \varepsilon \chi \iota
\kappa \psi \varsigma }=\dprod\limits_{\varkappa =\tau ,\upsilon ,\omega
,\varphi ,\rho ,\sigma ,\varepsilon ,\chi ,\iota ,\kappa ,\psi ,\varsigma
}(-e^{-t})^{\left\vert \varkappa \right\vert }q^{\kappa \left( \varkappa
\right) },
\end{equation}%
where $\kappa \left( \mu \right) $ is the second Casimir of the 2d partition
$\mu $ \textrm{(\ref{k}). }The factors $C_{\alpha \beta \gamma }$ are given
by eq(\ref{c})

\subsubsection{Partition function for the local 2-torus}

\qquad The partition function $Z_{H_{3}}$ of the local elliptic curve may be
obtained by implementing in $Z_{X_{4}}$ the constraint relations (\ref{st})
capturing the projection $X_{4}\rightarrow H_{3}$. \newline
Choosing trivial boundary conditions for the external legs and using:\newline
(\textbf{i}) the expression of the 4-vertex (\ref{cabc}), \newline
(\textbf{ii}) the rules of the \emph{planar} vertex formalism of \cite{9},%
\newline
we can write down directly the expression of the partition function $%
Z_{H_{3}}$. We find%
\begin{equation}
Z_{H_{3}}=\sum_{\xi ,\rho ,\sigma ,\eta ,\upsilon ,\tau ,\varsigma ,\theta
}\left( \mathcal{A}_{\omega \varphi ^{\mathrm{T}}\rho \sigma ^{\mathrm{T}%
}}^{\ast }\mathcal{B}_{\chi \rho ^{\mathrm{T}}\psi \varsigma ^{\mathrm{T}%
}}^{\ast }\mathcal{F}_{\iota \psi ^{\mathrm{T}}\kappa \omega ^{\mathrm{T}%
}}^{\ast }\mathcal{G}_{\varphi \kappa ^{\mathrm{T}}\chi \sigma ^{\mathrm{T}%
}\iota \varsigma ^{\mathrm{T}}}^{\ast }\mathcal{H}_{\omega \varphi \rho
\sigma \chi \iota \kappa \psi \varsigma }^{\ast }\right) ,
\end{equation}%
with
\begin{eqnarray}
\mathcal{A}_{\omega \varphi ^{\mathrm{T}}\rho \sigma ^{\mathrm{T}}}^{\ast }
&=&C_{\emptyset \omega \varphi ^{\mathrm{T}}}C_{\emptyset \rho \sigma ^{%
\mathrm{T}}},  \notag \\
\mathcal{B}_{\chi \rho ^{\mathrm{T}}\psi \varsigma ^{\mathrm{T}}}^{\ast }
&=&C_{\emptyset \chi \rho ^{\mathrm{T}}}C_{\emptyset \psi \varsigma ^{%
\mathrm{T}}},  \notag \\
\mathcal{F}_{\iota \psi ^{\mathrm{T}}\kappa \omega ^{\mathrm{T}}}^{\ast }
&=&C_{\emptyset \iota \psi ^{\mathrm{T}}}C_{\emptyset \kappa \omega ^{%
\mathrm{T}}},  \label{abcd} \\
\mathcal{G}_{\varphi \kappa ^{\mathrm{T}}\chi \sigma ^{\mathrm{T}}\iota
\varsigma ^{\mathrm{T}}}^{\ast } &=&C_{\emptyset \varphi \kappa ^{\mathrm{T}%
}}C_{\emptyset \chi \sigma ^{\mathrm{T}}}C_{\emptyset \iota \varsigma ^{%
\mathrm{T}}},,  \notag
\end{eqnarray}%
together with
\begin{equation}
\mathcal{H}_{\omega \varphi \rho \sigma \chi \iota \kappa \psi \varsigma
}^{\ast }=\dprod\limits_{\mu =\left\{ \omega ,\varphi ,\rho ,\sigma ,\chi
,\iota ,\kappa ,\psi ,\varsigma \right\} }(-e^{-t})^{\left\vert \mu
\right\vert }q^{\kappa \left( \mu \right) }  \label{ah}
\end{equation}%
where the factors $C_{\alpha \beta \gamma }$ are as in eq(\ref{c}).\newline
In what follows, we turn to study the field theory set up of the local
2-torus by starting by local $\mathbb{P}^{2}$ model.

\section{Sigma model for local $\mathbb{P}^{2}$}

\qquad In this section, we first review briefly the supersymmetric sigma
linear model realization of the local $\mathbb{P}^{2}$ model. This model is
useful for the purpose of this paper. \newline
We also use this field realization to fix convention notations and to
introduce some mathematical objects and their physical interpretations.

The local $\mathbb{P}^{2}$ model is nicely formulated in the language of $4D$%
, $\mathcal{N}=1$ supersymmetry which is, roughly, equivalent to the usual $%
2D$, $\mathcal{N}=2$ supersymmetry. The complex two dimension projective
plane $\mathbb{P}^{2}$ has one Kahler parameter $t$, interpreted as the
Fayet-Iliopoulos (FI) coupling constant in the supersymmetric gauge theory.%
\newline
The $U\left( 1\right) $ gauged linear sigma theory describing the local $%
\mathbb{P}^{2}$ target space geometry involves the following $4D$, $\mathcal{%
N}=1$ superfields (supersymmetric representations):\newline
(\textbf{1}) A $U\left( 1\right) $ gauge superfield $V=V\left( x,\theta ,%
\overline{\theta }\right) $ which reads, in the Wess-Zumino gauge, as
follows:
\begin{equation}
V=-\theta \sigma ^{\mu }\overline{\theta }A_{\mu }-i\overline{\theta }%
^{2}\theta \lambda +i\theta ^{2}\overline{\theta }\overline{\lambda }+\frac{1%
}{2}\theta ^{2}\overline{\theta }^{2}D,  \label{2}
\end{equation}%
where $\left( x^{\mu },\theta ^{a},\overline{\theta }_{\dot{a}}\right) $
stands for the $4D$, $\mathcal{N}=1$ superspace coordinates. \newline
In this relation, $A_{\mu }\left( x\right) $ and $\left( \lambda _{a}\left(
x\right) ,\overline{\lambda }_{\dot{a}}\left( x\right) \right) $ are
respectively the $U\left( 1\right) $ gauge vector and gaugino fields.
\newline
The scalar field $D$ is the usual auxiliary field capturing the local
Calabi-Yau geometry. It captures as well as part of the scalar field
potential $V$ of the gauge theory%
\begin{equation}
V=D^{2}+\sum_{i}\left\vert F_{i}\right\vert ^{2},
\end{equation}%
where the $F_{i}$ terms will be introduced below.\newline
(\textbf{2}) Four chiral superfields $\left\{ \Phi _{0},\Phi _{1},\Phi
_{2},\Phi _{3}\right\} $, with $\theta $-expansion
\begin{equation}
\Phi _{i}=z_{i}+\theta \psi _{i}+\theta ^{2}F_{i},  \label{3}
\end{equation}%
with $z_{i}$ being the field coordinates of local $\mathbb{P}^{2}$, $\psi
_{i}$ the Weyl spinors and $F_{i}$ the so called F-auxiliary fields. \newline
The $\Phi _{i}$ complex superfields carry the following $q_{i}$- charges
under the $U\left( 1\right) $ gauge symmetry,
\begin{equation}
\left( q_{0},q_{1},q_{2},q_{3}\right) =\left( -3,1,1,1\right) .  \label{4}
\end{equation}%
The $q_{i}$'s add exactly to zero as required by the Calabi-Yau condition
\begin{equation}
\sum_{i=0}^{3}q_{i}=0,  \label{5}
\end{equation}%
of local $\mathbb{P}^{2}$. \newline
The superfield Lagrangian density $\mathcal{L}_{\text{local }P^{2}}=\mathcal{%
L}\left( \Phi ,V\right) $ of the local $\mathbb{P}^{2}$ model reads, in the $%
\mathcal{N}=1$ $D=4$ formalism, as follows,
\begin{equation}
\mathcal{L}\left( \Phi ,V\right) =\int d^{4}\theta \sum_{i=0}^{3}\overline{%
\Phi }_{i}e^{2q_{i}V}\Phi _{i}-2t\int d^{4}\theta V+\mathcal{L}_{{\small %
gauge}}\left( V\right) .  \label{6}
\end{equation}%
Here $\mathcal{L}_{gauge}\left( V\right) $ is the superspace Lagrangian
density for the $U\left( 1\right) $ vector multiplet that can be found in
\textrm{\cite{18}}. $\newline
$The equation of motion of the auxiliary field D leads to,
\begin{equation}
\left\vert z_{1}\right\vert ^{2}+\left\vert z_{2}\right\vert ^{2}+\left\vert
z_{3}\right\vert ^{2}-3\left\vert z_{0}\right\vert ^{2}=t.  \label{7}
\end{equation}%
It is nothing but the defining equation of the local projective plane $%
\mathcal{O}\left( -3\right) \rightarrow \mathbb{P}^{2}$. \newline
The compact part $\mathbb{P}^{2}$ of this threefold is a complex plane given
by the divisor $z_{0}=0$; it is parameterized by the complex coordinates $%
\left( z_{1},z_{2},z_{3}\right) $ describing a complex surface embedded in $%
\mathbb{C}^{3}$ and has a $U\left( 1\right) $ gauge symmetry rotating the
phases of the coordinates variables. \newline
By setting $x_{i}=\left\vert z_{i}\right\vert ^{2}$, the complex surface $%
z_{0}=0$ can be represented by the planar triangle \cite{19},
\begin{equation}
x_{1}+x_{2}+x_{3}=t  \label{8}
\end{equation}%
with Kahler modulus $t$; see also figure 1. \newline
Because of the symmetry under permutation of the $x_{i}$'s, the triangle is
equilateral. The length of its edges are equal to $t$ and then it has an
area given by $A=\frac{t^{2}\sqrt{3}}{2}$.

\section{Field model for $\mathcal{O}\left( -3\right) \rightarrow E^{\left(
t,\infty \right) }$}

\qquad We first study the gauge invariant supersymmetric field model with
target space given by the curve $E^{\left( t,\infty \right) }=\partial
\mathbb{P}^{2}$. For details on the degenerate elliptic curve $E^{\left(
t,\infty \right) }$, see the appendix. Then, we consider the extension to $%
O\left( -3\right) \rightarrow E^{\left( t,\infty \right) }$.

\subsection{Divisors of local $\mathbb{P}^{2}$}

\qquad To begin, notice that the local $\mathbb{P}^{2}$ eq(\ref{7}) has
several divisors; i.e codimension one subspaces describing boundary patches
of the normal bundle of the projective plane. \newline
The standard ones are obtained by setting one of the $z_{i}$'s to zero; $%
z_{i}=0$ with $i=0,1,2,3$.

\subsubsection{Toric boundary of $\mathbb{P}^{2}$}

In this paragraph, we consider the three following complex surfaces $\left[
D_{1}\right] $, $\left[ D_{2}\right] $ and $\left[ D_{3}\right] $,
\begin{eqnarray}
\left[ D_{1}\right] &:&\left\vert z_{2}\right\vert ^{2}+\left\vert
z_{3}\right\vert ^{2}-3\left\vert z_{0}\right\vert ^{2}=t,\qquad
\Leftrightarrow \qquad z_{1}=0,  \notag \\
\left[ D_{2}\right] &:&\left\vert z_{3}\right\vert ^{2}+\left\vert
z_{1}\right\vert ^{2}-3\left\vert z_{0}\right\vert ^{2}=t,\qquad
\Leftrightarrow \qquad z_{2}=0,  \label{9} \\
\left[ D_{3}\right] &:&\left\vert z_{1}\right\vert ^{2}+\left\vert
z_{2}\right\vert ^{2}-3\left\vert z_{0}\right\vert ^{2}=t,\qquad
\Leftrightarrow \qquad z_{3}=0,  \notag
\end{eqnarray}%
and\textrm{\ }their union $\left[ D\right] =\left[ D_{1}\right] \cup \left[
D_{2}\right] \cup \left[ D_{3}\right] $. \newline
To see what this local geometry describes precisely; let us set $\left\vert
z_{0}\right\vert ^{2}=0$ in above eqs from where one sees that each relation
describes a complex one dimension projective line $\mathbb{P}^{1}$. To
distinguish between these complex projective lines, we use the convention
notation $\mathbb{P}_{i}^{1}$ where the subindex $i$ refers to $z_{i}=0$.
Thus we have
\begin{eqnarray}
\mathbb{P}_{1}^{1} &:&\left\vert z_{2}\right\vert ^{2}+\left\vert
z_{3}\right\vert ^{2}=t,  \notag \\
\mathbb{P}_{2}^{1} &:&\left\vert z_{3}\right\vert ^{2}+\left\vert
z_{1}\right\vert ^{2}=t,  \label{10} \\
\mathbb{P}_{3}^{1} &:&\left\vert z_{1}\right\vert ^{2}+\left\vert
z_{2}\right\vert ^{2}=t.  \notag
\end{eqnarray}%
As we see, these projective lines have the following intersection matrix%
\footnote{%
Denoting by $\left\{ \alpha _{1},\alpha _{2},\alpha _{3}\right\} $, a basis
of H$^{2}\left( P^{2},R\right) $, and by $\left\{ A^{1},A^{2},A^{3}\right\} $%
, the dual basis of $H_{2}\left( P^{2},R\right) $ with $\int_{A^{i}}\alpha
_{j}=\delta _{j}^{i}$, the intersection matrix eq(\ref{11}) is given by $%
I_{ij}=\int_{P^{2}}\alpha _{i}\wedge \alpha _{j}$.}
\begin{equation}
\mathbb{P}_{i}^{1}\cap \mathbb{P}_{j}^{1}=\left(
\begin{array}{ccc}
-2 & 1 & 1 \\
1 & -2 & 1 \\
1 & 1 & -2%
\end{array}%
\right) ,  \label{11}
\end{equation}%
from which one sees that the complex curve
\begin{equation}
E^{\left( t,\infty \right) }=\mathbb{P}_{1}^{1}\cup \mathbb{P}_{2}^{1}\cup
\mathbb{P}_{3}^{1},
\end{equation}%
is elliptic ($E^{\left( t,\infty \right) }\sim \mathbb{T}^{2}$). Indeed,
computing
\begin{equation}
E^{\left( t,\infty \right) }\cap E^{\left( t,\infty \right) }=\sum_{i=1}^{3}%
\mathbb{P}_{i}^{1}\cap \mathbb{P}_{i}^{1}+2\left( \mathbb{P}_{1}^{1}\cap
\mathbb{P}_{2}^{1}+\mathbb{P}_{2}^{1}\cap \mathbb{P}_{3}^{1}+\mathbb{P}%
_{3}^{1}\cap \mathbb{P}_{1}^{1}\right) ,
\end{equation}%
we get, up on using (\ref{11}),
\begin{equation}
E^{\left( t,\infty \right) }\cap E^{\left( t,\infty \right) }=-3\times
2+2\times 3=0.  \label{12}
\end{equation}%
From the toric diagram of $\mathbb{P}^{2}$, one also see that $E^{\left(
t,\infty \right) }$ describes indeed a toric complex one dimensional curve
defining the toric boundary of $\mathbb{P}^{2}$; i.e:
\begin{equation}
E^{\left( t,\infty \right) }\equiv \partial \mathbb{P}^{2}.  \label{13}
\end{equation}%
It is this curve that will be used to deal with the local 2-torus in the
large complex structure limit.

\subsubsection{Elliptic curve $E^{\left( t,\infty \right) }$}

\qquad An interesting question concerns the derivation of the defining
equation describing the elliptic curve $E^{\left( t,\infty \right) }$. From
the above analysis, it is not difficult to see that $E^{\left( t,\infty
\right) }$ is given by the following system of equations,
\begin{equation}
\left\{
\begin{array}{c}
\left\vert z_{1}\right\vert ^{2}+\left\vert z_{2}\right\vert ^{2}+\left\vert
z_{3}\right\vert ^{2}=t \\
z_{i}\equiv z_{i}e^{iq_{i}\alpha }, \\
z_{1}z_{2}z_{3}=0.%
\end{array}%
\right.  \label{14}
\end{equation}%
In these relations, we have three complex variables $\left(
z_{1},z_{2},z_{3}\right) $ subject to three constraint eqs,
\begin{figure}[tbph]
\begin{center}
\hspace{-1cm} \includegraphics[width=12cm]{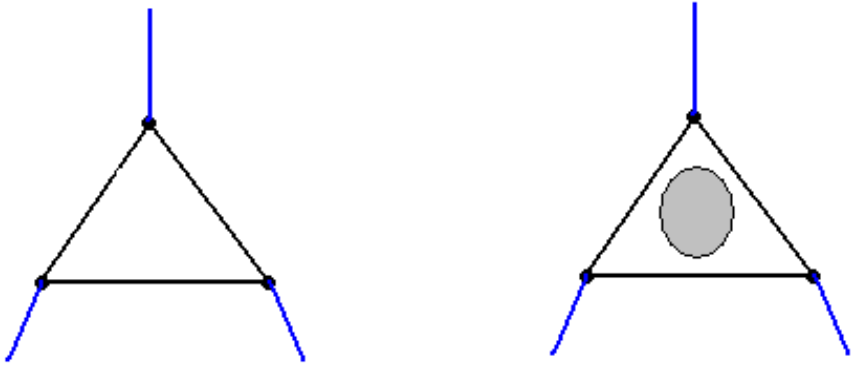}
\end{center}
\caption{(a) Left: {\protect\small {Toric graph of $\mathbb{P}^{2}$. (b)
Right: toric graph of E}}$^{\left( t,\infty \right) }${\protect\small {\
where we have added a hole to avoid confusion.}}}
\end{figure}
The two first eqs, which are real, are just the defining linear sigma model
eq of $\mathbb{P}^{2}$. They will be interpreted as the field equation of
motion of the \emph{auxiliary} \emph{D- field} in supersymmetric sigma model.%
\newline
The third relation, which is covariant under $U\left( 1\right) $ gauge
symmetry, is an extra complex condition implemented in order to restrict $%
\mathbb{P}^{2}$ geometry to its toric boundary $\partial \mathbb{P}^{2}$. It
will be interpreted later as the equation of motion of the \emph{auxiliary
F- fields}.\newline
Notice that, the implementation of the boundary condition is a new feature.
It can be then viewed as:\newline
(\textbf{i}) a generalization of the usual approach for dealing with sigma
model realization of toric manifolds. \newline
(\textbf{ii}) a way to approach genus g Riemann surfaces. \newline
(\textbf{iii}) a method that can be used to describe the toric boundary of
more general complex n dimensional toric Calabi-Yau manifolds. We will make
a comment regarding this point in the conclusion section. \newline
Notice finally that for $t\neq 0$ the three complex variables cannot vanish
simultaneously, i.e
\begin{equation}
\left( z_{1},z_{2},z_{3}\right) \neq \left( 0,0,0\right)  \label{15}
\end{equation}%
In the particular case $t=0$, the geometry collapses to the origin $\left(
0,0,0\right) $ where live a $\mathbb{P}^{2}$ singularity and an elliptic one.

\subsubsection{Divisor $\mathcal{O}\left( -3\right) \rightarrow E^{\left(
t,\infty \right) }$}

\qquad Using the above result on the toric realization of the elliptic
curve, one can immediately write down the defining equation of the divisor $%
\mathcal{O}\left( -3\right) \rightarrow E^{\left( t,\infty \right) }$ of the
local $\mathbb{P}^{2}$. We have
\begin{equation}
\left\{
\begin{array}{c}
\left\vert z_{1}\right\vert ^{2}+\left\vert z_{2}\right\vert ^{2}+\left\vert
z_{3}\right\vert ^{2}-3\left\vert z_{0}\right\vert ^{2}=t \\
z_{i}\equiv z_{i}e^{iq_{i}\alpha },\qquad i=0,1,2,3, \\
z_{1}z_{2}z_{3}=0,%
\end{array}%
\right.  \label{16}
\end{equation}%
where $\left( q_{0},q_{1},q_{2},q_{3}\right) $ are as in eq(\ref{4}) and
where the complex variable $z_{0}$ parameterizes the non compact direction $%
\mathcal{O}\left( -3\right) $. \newline
If we do not worry about the Calabi-Yau condition, the first relation can be
extended as
\begin{equation}
\left\vert z_{1}\right\vert ^{2}+\left\vert z_{2}\right\vert ^{2}+\left\vert
z_{3}\right\vert ^{2}-m\left\vert z_{0}\right\vert ^{2}=t.  \label{17}
\end{equation}%
where $m$ is an arbitrary positive integer.

\subsection{Superfield action}

\qquad Here we give the supersymmetric field description of (\ref{17}). We
start by studying the field realization of the toric curve $E^{\left(
t,\infty \right) }$. Then we consider its extension to the local geometry.

\subsubsection{Gauge invariant model for the elliptic curve}

\qquad To build the supersymmetric model describing the toric curve $%
E^{\left( t,\infty \right) }$, we start from the superfield content eq(\ref%
{7}) of local $\mathbb{P}^{2}$ theory and implement the constraint equation (%
\ref{16}) by using Lagrange multiplier method together $U\left( 1\right) $
gauge invariance. \newline
The appropriate Lagrange superfield multiplier is given by a chiral
superfield $\Upsilon $ with charge $q_{\Upsilon }=-3$ under $U\left(
1\right) $ gauge symmetry so that the chiral superfield monomial
\begin{equation}
W\left( \Phi ,\Upsilon \right) =\Phi _{1}\Phi _{2}\Phi _{3}\Upsilon ,
\label{18}
\end{equation}%
is gauge invariant. Thus the supersymmetric Lagrangian super-density with
target space $E^{\left( t,\infty \right) }$ is given by the density,
\begin{equation}
\mathcal{L}_{{\small E}}=\mathcal{L}_{P^{2}}+\left( g\int d^{2}\theta
W\left( \Phi ,\Upsilon \right) +hc\right) ,  \label{19}
\end{equation}%
where $g$ is a complex coupling constant. Since $\Upsilon $ has no kinetic
term, its elimination through the equation of motion
\begin{equation}
\frac{\delta \mathcal{L}_{{\small E}}}{\delta \Upsilon }=0  \label{20}
\end{equation}%
gives
\begin{equation}
g\Phi _{1}\Phi _{2}\Phi _{3}=0,  \label{21}
\end{equation}%
whose lowest term is precisely $z_{1}z_{2}z_{3}=0$. \newline
Notice that contrary to $\mathbb{P}^{2}$, the superfield realization of the
curve $E^{\left( t,\infty \right) }$ has a non trivial chiral
superpotential. As we see, this result is a particular situation that can be
extended to build toric realization of other toric manifolds. \newline
Notice also that the Lagrange superfield multiplier $\Upsilon $ can be given
a geometric interpretation. This superfield has no kinetic term $\overline{%
\Upsilon }\Upsilon $ nor couplings to the gauge superfield $V$; i.e no term
type
\begin{equation}
\int d^{4}\theta \overline{\Upsilon }e^{2qV}\Upsilon ,  \label{22}
\end{equation}%
in the Lagrangian super-density. The lack of (\ref{22}) can be interpreted
as corresponding to \emph{freezing} the supersymmetric gauge invariant
dynamics of $\Upsilon $. This property explains why the Calabi-Yau condition
for the complex toric curve $E^{\left( t,\infty \right) }$ should read as,
\begin{equation}
\sum_{i=1}^{3}q_{i}+q_{\gamma }=\sum_{i=1}^{3}q_{i}-3=0.  \label{23}
\end{equation}%
We will turn to this property when we consider the local threefold $\mathcal{%
O}\left( m\right) \oplus \mathcal{O}\left( -m\right) \rightarrow E^{\left(
t,\infty \right) }$. \newline
Notice also that the chiral superpotential (\ref{18}) is not the unique
gauge invariant term one may have. The general form of $W\left( \Phi
,\Upsilon \right) $ is given by
\begin{equation}
W\left( \Phi ,\Upsilon \right) =\sum_{n_{1}+n_{2}+n_{3}=3}\mathrm{g}_{%
{\small n}_{1}{\small ,n}_{2}{\small ,n}_{3}}\Phi _{1}^{n_{1}}\Phi
_{2}^{n_{2}}\Phi _{3}^{n_{3}}.
\end{equation}%
We will discuss this point in section 5 when we study the generalization to
higher dimension CY manifolds. \newline
For the moment, let us complete this discussion by giving the gauged
superfield realization of the complex surface $\mathcal{O}\left( -m\right)
\rightarrow E^{\left( t,\infty \right) }$.

\subsubsection{Field model for the Divisor $\mathcal{O}\left( -m\right)
\rightarrow E^{\left( t,\infty \right) }$}

\qquad In addition to the $U\left( 1\right) $ gauge superfield $V$, this
model involves five chiral superfields $\left( \Phi _{0},\Phi _{1},\Phi
_{2},\Phi _{3},\Upsilon \right) $ with charges
\begin{equation}
\left( q_{0},q_{1},q_{2},q_{3},q_{\gamma }\right) =\left( -m,1,1,1,-3\right)
.  \label{24}
\end{equation}%
The Lagrangian super-density $\mathcal{L}_{divisor}$ is given by,
\begin{eqnarray}
\mathcal{L}_{{\small divisor}} &=&\int d^{4}\theta \sum_{i=0}^{3}\overline{%
\Phi }_{i}e^{2q_{i}V}\Phi _{i}+\mathcal{L}_{{\small gauge}}\left( V\right)
-2t\int d^{4}\theta V  \notag \\
&&+\left( g\int d^{2}\theta W\left( \Phi ,\Upsilon \right) +hc\right)
\label{25}
\end{eqnarray}%
where the chiral superpotential $W\left( \Phi ,\Upsilon \right) $ is as in
eq(\ref{18}). Here also the first Chern class of the complex surface has a
contribution coming from $\Upsilon $ and reads as
\begin{equation}
\sum_{i=0}^{3}q_{i}+q_{\gamma }=-m
\end{equation}%
showing, as expected, that $\mathcal{O}\left( -m\right) \rightarrow
E^{\left( t,\infty \right) }$ is not a Calabi-Yau surface.

\subsection{Moduli space of supersymmetric vacuum}

\qquad Here we study the supersymmetric vacuum of the field model (\ref{25}%
). We show that the surface (\ref{16}) corresponds to a particular vacuum
given by the vev $z_{\gamma }=0$.

\textit{Moduli space of vacua}\newline
In the supersymmetric vacuum, the vanishing condition of the scalar
potential $V=V\left( z\right) $ of the model (\ref{19}) reads as
\begin{equation}
\left\vert D\right\vert ^{2}+\left\vert F_{0}\right\vert ^{2}+\left\vert
F_{1}\right\vert ^{2}+\left\vert F_{2}\right\vert ^{2}+\left\vert
F_{3}\right\vert ^{2}+\left\vert F_{\gamma }\right\vert ^{2}=0.  \label{26}
\end{equation}%
The dependence of the scalar potential $V$ in the scalar fields $z$ is
obtained by replacing the auxiliary fields $D$ and $F_{i}$ by their explicit
expressions in terms of the matter fields
\begin{equation}
\begin{tabular}{llll}
$D$ & $=$ & $D\left( z_{0},z_{1},z_{2},z_{3},z_{\gamma }\right) $ & , \\
$F_{i}$ & $=$ & $F_{i}\left( z_{0},z_{1},z_{2},z_{3},z_{\gamma }\right) $ & .%
\end{tabular}
\label{27}
\end{equation}%
These expressions are obtained by using the equations of motion
\begin{equation}
\frac{\delta \mathcal{L}}{\delta D}=0,\qquad \frac{\delta \mathcal{L}}{%
\delta \overline{F}_{i}}=0.
\end{equation}%
Eq(\ref{26}) is solved as follows,
\begin{equation}
D=0,\qquad F_{i}=0.
\end{equation}%
As noted before, $D=0$ leads to
\begin{equation}
\left\vert z_{1}\right\vert ^{2}+\left\vert z_{2}\right\vert ^{2}+\left\vert
z_{3}\right\vert ^{2}-m\left\vert z_{0}\right\vert ^{2}=t  \label{28}
\end{equation}%
and describes local $\mathbb{P}^{2}$ for the particular case $m=3$. \newline
$F_{i}=0$ involves five terms: $F_{0}$ which is trivial, and the remaining $%
F_{\gamma }$, $F_{3},$ $F_{2}$, and $F_{1}$ lead to:
\begin{eqnarray}
z_{1}z_{2}z_{3} &=&0,  \notag \\
z_{1}z_{2}z_{\gamma } &=&0,  \notag \\
z_{3}z_{1}z_{\gamma } &=&0,  \label{29} \\
z_{2}z_{3}z_{\gamma } &=&0,  \notag
\end{eqnarray}%
where $z_{\gamma }$\ stands for the lowest component field of the chiral
superfield $\Upsilon $. There are several solutions of these relations.
These solutions may be classified into two sets: \newline
(\textbf{1}) the first set is given by
\begin{equation}
z_{\gamma }=0.  \label{30}
\end{equation}%
Consequently eqs(\ref{29}) reduce to the first equation $z_{1}z_{2}z_{3}=0$.
\newline
The moduli background associated with this solution describes exactly the
complex surface $\mathcal{O}\left( -3\right) \rightarrow E^{\left( t,\infty
\right) }$.\newline
(\textbf{2}) the second set corresponds to
\begin{equation}
z_{\gamma }\neq 0
\end{equation}%
and two variables amongst the three $z_{1},$ $z_{2}$ and $z_{3}$ vanish.
\newline
So eqs(\ref{28}) become
\begin{eqnarray}
z_{2} &=&z_{3}=0:\qquad \left\vert z_{1}\right\vert ^{2}-m\left\vert
z_{0}\right\vert ^{2}=t,  \notag \\
z_{1} &=&z_{3}=0:\qquad \left\vert z_{2}\right\vert ^{2}-m\left\vert
z_{0}\right\vert ^{2}=t, \\
z_{1} &=&z_{2}=0:\qquad \left\vert z_{3}\right\vert ^{2}-m\left\vert
z_{0}\right\vert ^{2}=t.  \notag
\end{eqnarray}%
\emph{Case} $z_{\gamma }=0$ \newline
Let us now consider the interesting case $z_{\gamma }=0$ and study the
solution of constraint eq $z_{1}z_{2}z_{3}=0$. Here also there are several
solutions which we list below:\newline
(\textbf{1}) case $z_{3}=0$ but $\left( z_{1},z_{2}\right) \neq \left(
0,0\right) :\qquad $\newline
In this case the geometry reduces to
\begin{equation}
\left\vert z_{1}\right\vert ^{2}+\left\vert z_{2}\right\vert
^{2}-3\left\vert z_{0}\right\vert ^{2}=t.  \label{31}
\end{equation}%
It describes the local complex projective line
\begin{equation}
\mathcal{O}\left( -3\right) \rightarrow \mathbb{P}^{1},  \label{32}
\end{equation}%
which we denote as $\mathcal{O}\left( -3\right) \rightarrow \mathbb{P}%
_{3}^{1}$ where the sub-index 3 on $\mathbb{P}_{3}^{1}$ refers to $z_{3}=0$.
\newline
The same thing is valid for $z_{1}=0$ and $z_{2}=0$. \newline
They describe respectively the local surfaces $\mathcal{O}\left( -3\right)
\rightarrow \mathbb{P}_{1}^{1}$ and $\mathcal{O}\left( -3\right) \rightarrow
\mathbb{P}_{2}^{1}$.\newline
(\textbf{2}) case $z_{1}=0,$ $z_{2}=0,$ $z_{3}=\sqrt{t}$ \newline
This solution describes one of the three possible vertices of $\mathcal{O}%
\left( -3\right) \rightarrow E^{\left( t,\infty \right) }$; the two other
vertices are associated with the points:\newline
(i) $z_{1}=0,$ $z_{2}=\sqrt{t},$ $z_{3}=0$ and, \newline
(ii) $z_{1}=\sqrt{t},$ $z_{2}=0,$ $z_{3}=0$. \newline
(\textbf{3}) case $z_{1}=z_{2}=z_{3}=0$ is a $\mathbb{P}^{2}$ singularity.
\newline
This solution corresponds to the limit $t=0$ where both $E^{\left( t,\infty
\right) }$ and so $\mathbb{P}^{2}$ collapse down to a point.

The above analysis can be viewed as an interesting step towards the study of
topological vertex of local genus g- Riemann surfaces (in particular $g=1$)
by using toric diagrams based on the curve $E^{\left( t,\infty \right) }$.
To that purpose, one first has to build the toric realization of basic
objects of the topological vertex method. For instance, the complex
coordinates associated with the vertices of the elliptic curve $E^{\left(
t,\infty \right) }$ are given by the local patches
\begin{eqnarray}
\mathcal{U}_{1} &:&z_{0}^{\left( 1\right) },\quad z_{1}^{\left( 1\right)
},\quad z_{2}^{\left( 1\right) },\qquad z_{3}^{\left( 1\right) }=0,  \notag
\\
\mathcal{U}_{2} &:&z_{0}^{\left( 2\right) },\quad z_{1}^{\left( 2\right)
},\quad z_{3}^{\left( 2\right) },\qquad z_{2}^{\left( 2\right) }=0,
\label{33} \\
\mathcal{U}_{3} &:&z_{0}^{\left( 3\right) },\quad z_{2}^{\left( 3\right)
},\quad z_{3}^{\left( 3\right) },\qquad z_{1}^{\left( 3\right) }=0.  \notag
\end{eqnarray}%
The upper index of $z_{j}^{\left( i\right) }$ refers to the corresponding
chart $\mathcal{U}_{i}$. Note that on each chart, we have the relation
\begin{equation}
z_{1}^{\left( i\right) }z_{2}^{\left( i\right) }z_{3}^{\left( i\right)
}=0,\qquad i=1,2,3  \label{34}
\end{equation}%
The patches $\mathcal{U}_{i}$ could be interpreted as the three "toric
pants" needed for building $E^{\left( t,\infty \right) }$ and may be related
to the topological pant considered in \cite{11}. By gluing these three
patches, one reproduces $E^{\left( t,\infty \right) }$.

\section{Local 2-torus}

\qquad The local complex surface $X_{2}=\mathcal{O}\left( -m\right)
\rightarrow E^{\left( t,\infty \right) }$ we have considered so far is not a
Calabi-Yau 2-fold. The first Chern class $c_{1}\left( T^{\ast }X_{2}\right) $
of this variety is equal to $-m$. For our concern, this surface is thought
of as a divisor of the local Calabi-Yau threefold,
\begin{equation}
\mathcal{O}\left( m\right) \oplus \mathcal{O}\left( -m\right) \rightarrow
E^{\left( t,\infty \right) }.  \label{35}
\end{equation}

\subsection{Field model}

\begin{figure}[tbph]
\begin{center}
\hspace{-1cm} \includegraphics[width=14cm]{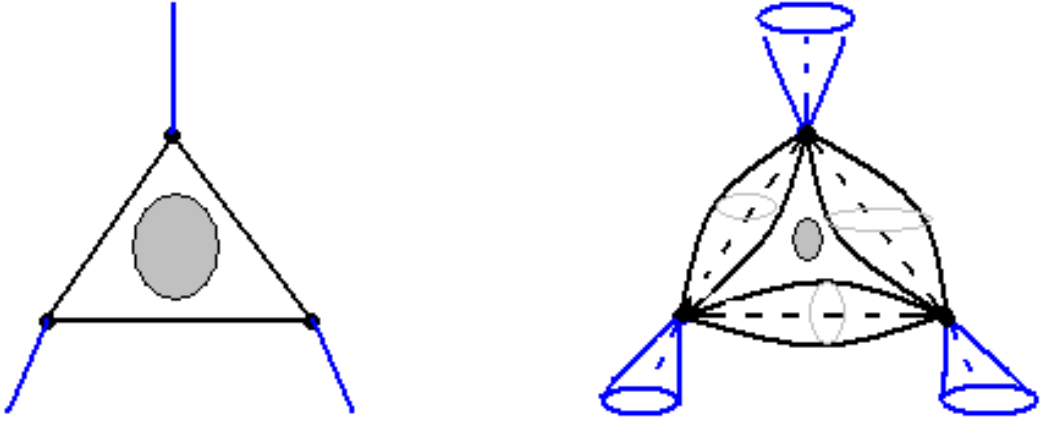}
\end{center}
\caption{{\protect\small {Toric graph of $\mathcal{O}\left( -3\right)
\rightarrow $}}$E^{\left( t,\infty \right) }${\protect\small {. The compact
part is }}$E^{\left( t,\infty \right) }${\protect\small {$=$}}$\left( {%
\partial \mathbb{P}^{2}}\right) ${\protect\small {\ with the usual three
vertices. Its toric frontier consists of three intersecting $\mathbb{P}^{1}$%
's in the homology class of 2-torus . (\textbf{a}) Figure (on left)
represents real skeleton. (\textbf{b}) Figure (on right) gives its fattening.%
}}}
\label{tr.eps}
\end{figure}

\qquad The supersymmetric field model relations describing the toric
Calabi-Yau threefold (\ref{35}) can be easily derived from previous study.
They are given by the following system of component field equations,
\begin{equation}
\left\{
\begin{array}{c}
\left\vert z_{1}\right\vert ^{2}+\left\vert z_{2}\right\vert ^{2}+\left\vert
z_{3}\right\vert ^{2}+3\left\vert z_{4}\right\vert ^{2}-3\left\vert
z_{0}\right\vert ^{2}=t, \\
z_{1}z_{2}z_{3}=0.%
\end{array}%
\right.  \label{36}
\end{equation}%
Here $z_{0}$ and $z_{4}$ parameterize the non compact directions and $\left(
z_{1},z_{2},z_{3}\right) $ are as before. The toric graph of this local
threefold is shown on \emph{figure }(\ref{tr.eps}); it has three
tetra-valent vertices,

To get the superfield Lagrangian density $\mathcal{L}_{locT^{2}}$, we think
about eqs(\ref{36}) as the field equations of motion of the $D$ and $F_{i}$
auxiliary fields. \newline
The first relation is associated with
\begin{equation}
\frac{\delta \mathcal{L}_{{\small locE}}}{\delta D}=0,  \label{37}
\end{equation}%
while the second follows from,
\begin{equation}
\frac{\delta \mathcal{L}_{{\small locE}}}{\delta F_{i}}=0.  \label{38}
\end{equation}%
The result is
\begin{eqnarray}
\mathcal{L}_{{\small locE}} &=&\int d^{4}\theta \sum_{i=1}^{3}\overline{\Phi
}_{i}e^{2V}\Phi _{i}+\int d^{4}\theta \left( \overline{\Phi }%
_{0}e^{-2mV}\Phi _{0}+\overline{\Phi }_{4}e^{2mV}\Phi _{4}\right)  \notag \\
&&+\mathcal{L}_{{\small gauge}}\left( V\right) -2t\int d^{4}\theta V+\left(
g\int d^{2}\theta \Phi _{1}\Phi _{2}\Phi _{3}\Upsilon +hc\right) ,
\label{39}
\end{eqnarray}%
where $\Upsilon $\ is a Lagrange superfield multiplier capturing the
constraint restricting the field variables to the boundary of $\mathbb{P}%
^{2} $.\newline
In addition to the $U\left( 1\right) $ gauge multiplet $V$, the chiral
superfields of this model are
\begin{equation}
\left( \Phi _{0},\Phi _{1},\Phi _{2},\Phi _{4},\Phi _{5},\Upsilon \right)
\label{40}
\end{equation}%
and carry the following $q_{i}$- charges under the $U\left( 1\right) $ gauge
symmetry,
\begin{equation}
\left( q_{0},q_{1},q_{2},q_{3},q_{4},q_{\gamma }\right) =\left(
-m,1,1,1,m,-3\right) .  \label{41}
\end{equation}%
where $m${\Huge \ }is a priori equal to{\Huge \ }$3$; but in general can
take any integral value.

\subsection{Generalization}

\qquad The construction we have developed here above can be generalized to
other Calabi-Yau manifolds. Below we make a comment on two kinds of
generalizations. The first extension deals with the gauged sigma model
realization of local genus g-Riemann surfaces for $g\geq 2$. The second
generalization concerns sigma model approach for higher complex dimensional
compact toric Calabi-Yau manifolds.

\subsubsection{Local genus g-Riemann surfaces}

\qquad So far we have seen that for each local elliptic curve, it is
associated a $U\left( 1\right) $ gauge symmetry. This gauge symmetry is
inherited from the $\mathbb{P}^{2}$ model. Since local genus g-Riemann
surfaces can be engineered by gluing several local elliptic curves, we
conclude that a class of local genus g-Riemann surfaces could be described
by higher rank abelian $U^{n}\left( 1\right) $ gauged supersymmetric field
model type. The rank $n$ of the gauge symmetry depends on the way the gluing
is done. \newline
To illustrate the idea, let us give the example of $g=2$- Riemann surface
described by a 2D $U^{2}\left( 1\right) $ gauged $\mathcal{N}=2$
supersymmetric sigma model. \newline
The local $g=2$- Riemann surface in the large complex structures limits can
be engineered by gluing two local elliptic curves with compact base $%
E_{1}^{\left( t,\infty \right) }=\partial \mathbb{P}_{1}^{2}$ and $%
E_{2}^{\left( t,\infty \right) }=\partial \mathbb{P}_{2}^{2}$. In the sigma
model approach, we distinguish different representations according to
whether $\mathbb{P}_{1}^{2}$ and $\mathbb{P}_{2}^{2}$ have an intersection
point or edge. \newline
In the first case, the sigma model involves five complex field variables,
\begin{equation}
\left( z_{1},z_{2},z_{3},z_{4},z_{5}\right)  \label{42}
\end{equation}%
and a $U^{2}\left( 1\right) $ gauge invariance under which these complex
variables have the following gauge charges
\begin{eqnarray}
\left( q_{i}^{1}\right) &=&\left( 1,1,1,0,0\right)  \notag \\
\left( q_{i}^{2}\right) &=&\left( 0,0,1,1,1\right)  \label{43}
\end{eqnarray}%
The field theoretic equations describing the compact part of the moduli
space of supersymmetric vacua are given by,
\begin{equation}
\left\{
\begin{array}{c}
\left\vert z_{1}\right\vert ^{2}+\left\vert z_{2}\right\vert ^{2}+\left\vert
z_{3}\right\vert ^{2}=t_{1} \\
z_{1}z_{2}z_{3}=0%
\end{array}%
\right. \qquad ,\qquad \left\{
\begin{array}{c}
\left\vert z_{3}\right\vert ^{2}+\left\vert z_{4}\right\vert ^{2}+\left\vert
z_{5}\right\vert ^{2}=t_{2} \\
z_{3}z_{4}z_{5}=0%
\end{array}%
\right. ,  \label{44}
\end{equation}%
where $t_{1}$ and $t_{2}$\ are respectively the Kahler moduli of the
projective planes $\mathbb{P}_{1}^{2}$ and $\mathbb{P}_{2}^{2}$. The
holomorphic constraint eqs $z_{1}z_{2}z_{3}=0$ and $z_{3}z_{4}z_{5}=0$ are
implemented in the gauged supersymmetric superfield model by two chiral
superfields $\Upsilon _{1}$ and $\Upsilon _{2}$ with gauge charges $\left(
q_{\gamma }^{1},q_{\gamma }^{2}\right) $ equal to $\left( -3,0\right) $ and $%
\left( 0,-3\right) $ respectively. \newline
Notice that\textrm{\ }eq(\ref{44}) describes indeed a complex curve with
genus $g=2$. The toric threefold based on this genus $g=2$ curve is
parameterized by seven complex variables,
\begin{equation}
\left( z_{0},z_{1},z_{2},z_{3},z_{4},z_{5},z_{6}\right) ,  \label{45}
\end{equation}%
with gauge charges as
\begin{eqnarray}
\left( q_{i}^{1}\right) &=&\left( -3,1,1,1,0,0,-3\right) ,  \notag \\
\left( q_{i}^{2}\right) &=&\left( -3,0,0,1,1,1,-3\right) .  \label{46}
\end{eqnarray}%
The gauged supersymmetric field theoretical equation
\begin{eqnarray}
-m\left\vert z_{0}\right\vert ^{2}+\left\vert z_{1}\right\vert
^{2}+\left\vert z_{2}\right\vert ^{2}+\left\vert z_{3}\right\vert
^{2}+m\left\vert z_{6}\right\vert ^{2} &=&t_{1},\qquad z_{1}z_{2}z_{3}=0,
\notag \\
-n\left\vert z_{0}\right\vert ^{2}+\left\vert z_{3}\right\vert
^{2}+\left\vert z_{4}\right\vert ^{2}+\left\vert z_{5}\right\vert
^{2}+n\left\vert z_{6}\right\vert ^{2} &=&t_{2},\qquad z_{3}z_{4}z_{5}=0,
\label{47}
\end{eqnarray}%
where $m$ and $n$ are in general arbitrary integers; but can be set equal to
$3$ to keep in touch with the first Chern class of the complex two
dimensional projective plane. \newline
These relations involve seven complex variables constrained by four complex
constraint eqs leaving then three complex variables free. Note also that the
first relation of the above equation describes $\mathcal{O}\left( m\right)
\oplus \mathcal{O}\left( -m\right) \rightarrow E_{1}^{\left( t,\infty
\right) }$ while the second describes $\mathcal{O}\left( n\right) \oplus
\mathcal{O}\left( -n\right) \rightarrow E_{2}^{\left( t,\infty \right) }$.

\subsubsection{Higher dimensional toric CY manifolds}

\qquad The gauged supersymmetric sigma model for the boundary surface of
local $\mathbb{P}^{2}$ that we have considered in this paper can be extended
for compact divisors of local $\mathbb{P}^{n-1}$. The latter is given by the
following $U\left( 1\right) $ gauge invariant complex dimension $n$
hypersurface
\begin{equation}
n\left\vert z_{0}\right\vert ^{2}+\sum_{i=1}^{n}\left\vert z_{i}\right\vert
^{2}=t,  \label{48}
\end{equation}%
embedded in $\mathbb{C}^{n+1}$ parameterized by the local coordinates $%
\left\{ z_{0},z_{1},z_{2},...,z_{n}\right\} $ with gauge charge
\begin{equation}
\left( q_{0},q_{1},q_{2},...,q_{n}\right) =\left( -n,1,1,...,1\right)
\label{49}
\end{equation}%
In eq(\ref{48}), $t$ is the usual Kahler parameter of $\mathbb{P}^{n-1}$. To
describe the compact (divisor) boundary $\partial \left( \mathbb{P}%
^{n-1}\right) $ of the toric n-fold, we supplement the hypersurface equation
by the following extra gauge covariant constraint relation,
\begin{equation}
\dprod\limits_{i=1}^{n}z_{i}=0.  \label{50}
\end{equation}%
Extending the analysis of section 4, the $U\left( 1\right) $ gauged
supersymmetric sigma model describing $\partial \left( \mathbb{P}%
^{n-1}\right) $ reads as follows
\begin{equation}
\mathcal{L}_{\partial P^{n-1}}=\mathcal{L}_{P^{n-1}}+\left( g\int
d^{2}\theta W\left( \Phi ,\Upsilon \right) +hc\right)  \label{51}
\end{equation}%
with chiral superpotential
\begin{equation}
W\left( \Phi ,\Upsilon \right) =\Upsilon \dprod\limits_{i=1}^{n}\Phi _{i}.
\label{52}
\end{equation}%
The gauge charge $q_{\gamma }$ of the Lagrange multiplier superfield $%
\Upsilon $ is equal to $\left( -n\right) $. Here also the first Chern class
of $\partial \left( \mathbb{P}^{n-1}\right) $ is identically zero. As noted
before, this property is not a new feature since the most general gauge
invariant chiral superpotential extending eq(\ref{52}) is given by
\begin{equation}
W\left( \Phi ,\Upsilon \right) =\sum_{m_{1}+...+m_{n}=n}g_{\left\{
m_{i}\right\} }\left( \Upsilon \dprod\limits_{i=1}^{n}\Phi
_{i}^{m_{i}}\right)  \label{53}
\end{equation}%
where $g_{\left\{ m_{i}\right\} }$ are complex coupling constants. \newline
The equation of motion of $\Upsilon $\ gives a degree $n$ homogeneous
polynom describing a complex $\left( n-2\right) $ dimension holomorphic CY
hypersurface with complex structures $g_{\left\{ m_{i}\right\} }$.

\section{Conclusion}

\qquad In this paper, we have set up the basis of the non planar topological
3-vertex method to compute the topological string amplitudes for the family
of local elliptic curves
\begin{equation}
\begin{tabular}{llllll}
$\mathcal{O}(m)\oplus \mathcal{O}(-m)$ & $\rightarrow $ & $E^{\left( t,%
\mathrm{\mu }\right) }$ & , & $m\in Z$ & ,%
\end{tabular}%
\end{equation}%
in the limit of large complex structure $\mathrm{\mu }$; i.e
\begin{equation}
\begin{tabular}{llll}
$\left\vert \mathrm{\mu }\right\vert $ & $\rightarrow $ & $\infty $ & .%
\end{tabular}%
\end{equation}%
Generally speaking, the base $E^{\left( t,\mathrm{\mu }\right) }$ stands for
an elliptic curve with Kahler parameter $t$ and complex structure $\mathrm{%
\mu }$ embedded in the projective plane $\mathbb{P}^{2}$. In the large limit
$\mathrm{\mu }$; the corresponding elliptic curve $E^{\left( t,\mathrm{%
\infty }\right) }$ is realized as the toric boundary of $\mathbb{P}^{2}$;
see appendix for more details; in particular eqs(\ref{t},\ref{mm},\ref{lcs}%
). \newline
First, we have reviewed the main idea of the usual (planar) topological
3-vertex method for non compact toric threefolds. \newline
Then, we have drawn the first lines of the non planar topological 3-vertex
method for the local degenerate 2-torus. The latter is a non compact toric
Calabi-Yau threefold given by a hypersurface in a complex Kahler 4-fold.
\newline
The key idea in getting the particular toric representation of the local
2-torus with large complex structure is based on thinking about $E^{\left(
t,\infty \right) }$ as given by the toric boundary of the complex projective
plane $\mathbb{P}^{2}$. In this view, $\mathcal{O}(m)\oplus \mathcal{O}%
(-m)\rightarrow E^{\left( t,\infty \right) }$ becomes a toric threefold and
so one may extend the results of the topological 3- vertex method of \cite{9}
to the case of the local (degenerate) 2-torus. Obviously, to compute the
topological amplitudes, we have to use the \emph{non planar} 3-vertex method
rather than the usual \emph{planar} 3-vertex one. Regarding this matter, we
have given first results concerning the local degenerate elliptic curve $%
\mathcal{O}(m)\oplus \mathcal{O}(-m)\rightarrow E^{\left( t,\infty \right) }$%
. More analysis is however still needed before getting the complete explicit
results.

We have also developed the gauged supersymmetric sigma model realization
that underlies the geometry the local 2-torus with $\left\vert \mathrm{\mu }%
\right\vert \longrightarrow \infty $ and exhibited explicitly the role of D-
and F- terms. We have discussed as well how this construction could be
extended to local genus g-Riemann surfaces $\mathcal{O}(m)\oplus \mathcal{O}%
(2-2g-m)$ $\rightarrow \Sigma _{g}$ in the limit of large complex
structures. \newline
The results obtained in the field theory part of the paper may also be
viewed as an explicit analysis regarding implementation of F- terms in the
Witten's original work on phases of $\mathcal{N}=2$ supersymmetric theories
in two dimensions\emph{\ }\cite{18}.

\begin{acknowledgement}
{\small \ The authors thank the International Centre for Theoretical
Physics, and S. Randjabar Daemi for kind hospitality at ICTP. This research
work is supported by Protars III CNRST-D12/25.}
\end{acknowledgement}

\section{Appendix}

In this appendix, we give useful properties on the complex projective plane
and on particular aspects on the complex curves in $\mathbb{P}^{2}$.\newline
More precisely denoting by $\mathbb{P}_{t}^{2}$, the projective plane with
Kahler parameter $t$ and by $E^{\left( t,\mu \right) }$ the following
elliptic curve in $\mathbb{P}_{t}^{2}$,%
\begin{equation*}
E^{\left( t,\mu \right) }:\qquad z_{1}^{3}+z_{2}^{3}+z_{3}^{3}+\mu
z_{1}z_{2}z_{3}=0
\end{equation*}%
we want to show that the boundary $\partial \left( \mathbb{P}_{t}^{2}\right)
$ is nothing but the degenerate limit $\mathrm{\mu \longrightarrow \infty }$
of $E^{\left( t,\mu \right) }$; that is
\begin{equation}
\partial \left( \mathbb{P}_{t}^{2}\right) \simeq E^{\left( t,\infty \right)
}.  \label{que}
\end{equation}%
This question can be also rephrased in other words by using the fibration,%
\begin{equation}
\mathbb{P}^{2}=\mathcal{B}_{2}\times \mathbb{T}^{2},  \label{fib}
\end{equation}%
where $\mathcal{B}_{2}$ is real 2 dimensional base (an equilateral
triangle). The boundary $\partial \left( \mathbb{P}^{2}\right) $ is a toric
submanifold with fibration
\begin{equation}
\begin{tabular}{llll}
$\partial \left( \mathbb{P}^{2}\right) $ & $=$ & $\Delta _{1}\times \mathbb{S%
}^{1}$ & ,%
\end{tabular}
\label{dp2}
\end{equation}%
where $\Delta _{1}=\left( \partial \mathcal{B}_{2}\right) $ is the boundary
of a triangle.\newline
Clearly, thought not exactly the standard 2- torus $\mathbb{S}^{1}\times
\mathbb{S}^{1}$, the boundary $\partial \left( \mathbb{P}^{2}\right) $ has
something to do with it. It is the large complex structure $\mu $ of the
elliptic curve $E^{\left( t,\mu \right) }$; say
\begin{equation}
\begin{tabular}{llll}
$\left\vert \mu \right\vert $ & $\longrightarrow $ & $+\infty $ & .%
\end{tabular}
\label{lim}
\end{equation}%
As we need both Kahler and complex structures to answer the question (\ref%
{que}), let us first give some useful details and then turn to derive the
identity $\partial \left( \mathbb{P}_{t}^{2}\right) \simeq E^{\left(
t,\infty \right) }$.\newline

\emph{Projective plane }$\mathbb{P}^{2}$\newline
There are different ways to deal the complex projective plane $\mathbb{P}%
^{2} $. Below, we give two dual descriptions by using the so called type IIA
and type IIB geometries \textrm{\cite{ge}}.

\emph{Type IIA geometry}\newline
In this set up, known also as toric geometry, the projective plane $\mathbb{P%
}^{2}$ is defined by the following real 4 dimensional compact hypersurface
in $\mathbb{C}^{3}$,%
\begin{equation}
\left\vert z_{1}\right\vert ^{2}+\left\vert z_{2}\right\vert ^{2}+\left\vert
z_{3}\right\vert ^{2}=t,  \label{t}
\end{equation}%
where $\left( z_{1},z_{2},z_{3}\right) $ stand for local complex
coordinates. In the above relation, the complex variables obey the gauge
identifications
\begin{equation}
\begin{tabular}{llll}
$z_{k}^{\prime }$ & $\equiv $ & $e^{i\varphi }z_{k}$ & ,%
\end{tabular}%
\end{equation}%
where the real phase $\varphi $ is the parameter of the $U\left( 1\right) $
gauge symmetry. The phase $\varphi $ can be used to fix one of the three
phases of the $z_{k}=\left\vert z_{k}\right\vert e^{i\varphi _{k}}$ complex
coordinates leaving then two free phases; say $\varphi _{1}$ and $\varphi
_{2}$. These free phases are precisely the ones used to parameterize the
2-torus in the fibration (\ref{fib}). \newline
The positive parameter $t$ is the Kahler modulus of the projective plane; it
controls the size of $\mathbb{P}^{2}$. Indeed, in the singular limit $%
t\longrightarrow 0$, we have the two following:\newline
(i) the size of the complex surface $\mathbb{P}^{2}$ vanishes%
\begin{equation}
\lim_{t\longrightarrow 0}\left[ vol\left( \mathbb{P}^{2}\right) \right] =0,
\end{equation}%
in agreement with both the relation $vol\left( \mathbb{P}^{2}\right) \sim
t^{2}$ (\emph{footnote 7}) and eq(\ref{t}) which becomes then singular.%
\newline
(ii) the size of the complex boundary $\partial \left( \mathbb{P}^{2}\right)
$ vanishes as well
\begin{equation}
\lim_{t\longrightarrow 0}\left[ vol\left[ \partial \left( \mathbb{P}%
^{2}\right) \right] \right] =0,
\end{equation}%
The two above relations show that the Kahler parameter $t$ of $\mathbb{P}%
^{2} $ and the Kahler parameter $r$\ of it boundary $\partial \left( \mathbb{%
P}^{2}\right) $ are intimately related. We will show later that they are the
same\textrm{\footnote{%
From eq(\ref{t}), it is not difficult to see that the volume of $\mathbb{P}%
^{2}$ is proportional to $t^{2}$ while the volume of its boundary $\partial
\left( \mathbb{P}^{2}\right) $ is proportional $t$.}}.\newline
Notice in passing that in the field theory language, the relation (\ref{t})
has an interpretation as the field equation of motion of the D- auxiliary
field in the $U\left( 1\right) $ gauged sigma model realization of $\mathbb{P%
}^{2}$. There, the Kahler parameter $t$ is interpreted as the
Fayet-Iliopoulos coupling constant term. This description is well known;
some of its basic aspects have been studied in section 4 of this the paper;
we will then omit redundant details.

\emph{Type IIB geometry}:\newline
In the type IIB geometry, one thinks about the complex projective surface $%
\mathbb{P}^{2}$ as a complex holomorphic algebraic surface obtained by
taking the coset of the complex space $\mathbb{C}^{3}\backslash \left\{
\left( 0,0,0\right) \right\} $ by the complex abelian group $\mathbb{C}%
^{\ast }$;%
\begin{equation}
\mathbb{P}^{2}=\left[ \mathbb{C}^{3}\backslash \left\{ \left( 0,0,0\right)
\right\} \right] /\mathbb{C}^{\ast }.
\end{equation}%
The $\mathbb{C}^{\ast }$ group action allows to make the following
identifications,%
\begin{equation}
\left( z_{1},z_{2},z_{3}\right) \equiv \left( \lambda z_{1},\lambda
z_{2},\lambda z_{3}\right)  \label{ctr}
\end{equation}%
with $\lambda $ being an arbitrary non zero complex number. This
identification reduces the complex 3- dimension down to complex \emph{2}
dimensions.\emph{\ }Here also, one can make gauge choices by working in a
particular local coordinate patch. A standard gauge choice is the one given
by the condition $\lambda z_{3}=1$.\newline

\emph{Complex curves in }$\mathbb{P}^{2}$\newline
Complex curves ( real Riemann surfaces) in $\mathbb{P}^{2}$ are complex
codimension one submanifolds obtained by imposing one more complex
constraint relation $\mathrm{f}\left( z_{i}\right) =0$ on the projective
complex variables $z_{1},z_{2}$ and $z_{3}$. The most common curves in $%
\mathbb{P}^{2}$ are obviously the projective lines $\mathbb{P}^{1}$, conics
and elliptic curves. \newline
Generally speaking, the constraint eq $\mathrm{f}\left( z_{i}\right) =0$ can
be stated as follows,
\begin{equation}
\mathrm{f}\left( \lambda z_{1},\lambda z_{2},\lambda z_{3}\right) =\lambda
^{n}\mathrm{f}\left( z_{1},z_{2},z_{3}\right) =0,  \label{f}
\end{equation}%
where $n$ stands for the degree of homogeneity of the curve.\ The case $n=3$%
, is given by the following typical cubic%
\begin{equation}
z_{1}^{3}+z_{2}^{3}+z_{3}^{3}+\mu z_{1}z_{2}z_{3}=0.  \label{mm}
\end{equation}%
This relation describes an elliptic curve $E$ of \emph{degree 3} with a
complex structure $\mu $. This curve $E$ has been extensively used in
physical literature; in particular in the geometric engineering of \emph{4D}
superconformal field theories embedded in \emph{10D}\ type IIB superstring
on elliptically fibered Calabi-Yau threefolds \textrm{\cite{ge,gf}}. \newline
Before proceeding further, it is interesting to notice that the elliptic
curve $E$ is a genus one Riemann surface having a real \emph{3d} moduli
space; parameterized by
\begin{equation}
\left( \mu _{1},\mu _{2};r\right)
\end{equation}%
with $\mu =\mu _{1}+i\mu _{2}$ is the complex structure and $r$ is its
Kahler modulus. So, elliptic curves may be generally denoted as follows
\begin{equation}
E^{\left( r,\mu \right) }.
\end{equation}%
Regarding the complex parameter $\mu $ of the elliptic curve $E^{\left(
r,\mu \right) }$, it is explicitly exhibited in type IIB geometry set up as
shown on eq(\ref{mm}). \newline
However, it is interesting to notice that the Kahler parameter $r$ cannot be
exhibited explicitly since $E^{\left( r,\mu \right) }$ has no standard type
IIA geometry realisation\textrm{\footnote{%
In toric geometry the real base of the fibration $\mathcal{B}_{n}\times
\mathbb{T}^{n}$ of complex n- dimensional toric manifolds involves
projective lines. Torii appear rather in the fiber.}} of the type given by
eq(\ref{t}).\newline
The construction developed in this paper gives a way to circumvent this
difficulty by using the degenerate representation (\ref{dp2}). \newline
With the above features in mind, we turn now to the derivation of eq(\ref%
{que}).

$\partial \left( \mathbb{P}^{2}\right) $\emph{\ as the degenerate elliptic
curve }$E^{\left( r,\infty \right) }$\emph{\ }\newline
Here we would like to show that $\partial \left( \mathbb{P}^{2}\right) $ is $%
E^{\left( r,\mu \right) }$ but with a large complex structure $\mu $; that
is $\left\vert \mu \right\vert \longrightarrow \infty $.\newline
To get the key point behind the identity (\ref{que}) as well as the
degeneracy of the elliptic curve $E^{\left( r,\mu \right) }$, we give the
two following properties:\newline
(\textbf{a}) the projective plane $\mathbb{P}^{2}$ has three particular
intersecting divisors $\mathcal{D}_{i}$. These are associated with the
hyperlines
\begin{equation}
z_{i}=0
\end{equation}%
in $\mathbb{P}^{2}$ which, up on using eq(\ref{t}), lead to the following
relations%
\begin{equation}
\begin{tabular}{llllll}
$\mathcal{D}_{1}$ & $:$ & $\left\vert z_{2}\right\vert ^{2}+\left\vert
z_{3}\right\vert ^{2}$ & $=$ & $t$ & $,$ \\
$\mathcal{D}_{2}$ & $:$ & $\left\vert z_{1}\right\vert ^{2}+\left\vert
z_{3}\right\vert ^{2}$ & $=$ & $t$ & $,$ \\
$\mathcal{D}_{3}$ & $:$ & $\left\vert z_{1}\right\vert ^{2}+\left\vert
z_{2}\right\vert ^{2}$ & $=$ & $t$ & $.$%
\end{tabular}%
\end{equation}%
From these equations, we see that each divisor $\mathcal{D}_{i}$ is a
projective line with Kahler parameter $t$. \newline
The equality of the Kahler parameters $t_{1}=t_{2}=t_{3}=t$ of these
projective lines may be also interpreted as due to the permutation symmetry
of the $\left\{ z_{i}\right\} $ projective coordinate variables of $\mathbb{P%
}^{2}$. This feature translates, in the language of toric geometry, as
corresponding to having an equilateral triangle for the real base $\mathcal{B%
}_{2}$.\newline
(\textbf{b}) The divisors $\left\{ \mathcal{D}_{i}\right\} $ are precisely
the ones we get by taking the large complex structure limit (\ref{lim}) of
the complex curve eq(\ref{mm}). Under this condition, eq(\ref{mm}) reduces
then to the dominant monomial
\begin{equation}
\mu z_{1}z_{2}z_{3}=0.  \label{lcs}
\end{equation}%
Notice that the above relation is obviously invariant under the $\mathbb{C}%
^{\ast }$ transformations (\ref{ctr}) since
\begin{equation}
\mu \left( \lambda z_{1}\right) \left( \lambda z_{2}\right) \left( \lambda
z_{3}\right) =\lambda ^{3}\left( \mu z_{1}z_{2}z_{3}\right) =0.
\end{equation}%
To have more insight about the elliptic curve $E^{\left( t,\infty \right) }$
with large complex structure; $\left\vert \mu \right\vert \longrightarrow
\infty $, it is interesting to solve eq(\ref{lcs}). There are three
solutions classified as follows:\newline
\textbf{(i)} $z_{1}=0$, what ever the two other complex variables $z_{2}$
and $z_{3}$ are; provided that
\begin{equation}
\begin{tabular}{llll}
$\left( z_{2},z_{3}\right) $ & $\neq $ & $\left( 0,0\right) $ & , \\
$\left( z_{2},z_{3}\right) $ & $\equiv $ & $\left( \lambda z_{2},\lambda
z_{3}\right) $ & .%
\end{tabular}%
\end{equation}%
But these relations are nothing but the definition of the divisor $\mathcal{D%
}_{1}$ in type IIB geometry.\newline
\textbf{(ii)} $z_{2}=0$, what ever the other complex variables $z_{1}$ and $%
z_{3}$ are; provided that
\begin{equation}
\begin{tabular}{llll}
$\left( z_{1},z_{3}\right) $ & $\neq $ & $\left( 0,0\right) $ & , \\
$\left( z_{1},z_{3}\right) $ & $\equiv $ & $\left( \lambda z_{1},\lambda
z_{3}\right) $ & ,%
\end{tabular}%
\end{equation}%
describing then the divisor $\mathcal{D}_{2}$.\newline
\textbf{(iii)} $z_{3}=0$, what ever the other complex variables $z_{2}$ and $%
z_{1}$ are; provided that
\begin{equation}
\begin{tabular}{llll}
$\left( z_{1},z_{2}\right) $ & $\neq $ & $\left( 0,0\right) $ & , \\
$\left( z_{1},z_{2}\right) $ & $\equiv $ & $\left( \lambda z_{1},\lambda
z_{2}\right) $ & ,%
\end{tabular}%
\end{equation}%
associated with the divisor $\mathcal{D}_{3}$.\newline
To conclude the boundary $\left( \partial \mathbb{P}_{t}^{2}\right) $ of the
projective plane $\mathbb{P}_{t}^{2}$ is indeed described by an elliptic
curve with a Kahler parameter $t$ inherited from the $\mathbb{P}_{t}^{2}$
one; but with a large complex structure $\mu $; see also \emph{footnote 7}.
The limit $\mu \rightarrow \infty $ explains the degeneracy property in the
base $\Delta _{1}$ (\ref{dp2}).


\begin{thebibliography}{99}
\bibitem{01} Marcos Marino, \emph{Chern-Simons Theory and Topological Strings%
}, Rev.Mod.Phys. 77 (2005) 675-720, arXiv:hep-th/0406005

\bibitem{02} R. Dijkgraaf, E. Verlinde, H. Verlinde, \textquotedblleft \emph{%
Notes on topological string theory and two-dimensional topological
gravity,\textquotedblright\ in String theory and quantum gravity}, World
Scientific Publishing, p. 91, (1991)

\bibitem{03} M. Bershadsky, S. Cecotti, H. Ooguri and C. Vafa, \emph{%
\textquotedblleft Kodaira-Spencer theory of gravity and exact results for
quantum string amplitudes,\textquotedblright } Commun. Math. Phys. 165, 311
(1994), hep-th/9309140.

\bibitem{1} G. L. Cardoso, B. de Wit, J. Kappeli, T. Mohaupt, \emph{%
Stationary BPS Solutions in N=2 Supergravity with R}$^{\emph{2}}$\emph{%
-Interactions}, JHEP 0012 (2000), 019arXiv:hep-th/0009234,

\bibitem{2} A. Ceresole, R. D'Auria, S. Ferrara, \emph{The Symplectic
Structure of N=2 Supergravity and its Central Extension},
Nucl.Phys.Proc.Suppl. 46 (1996) 67-74, arXiv:hep-th/9509160,

\bibitem{201} T. Graber and E. Zaslow, \emph{\textquotedblleft Open string
Gromov-Witten invariants: Calculations and a mirror
'theorem'\textquotedblright } arXiv:hep-th/0109075.

\bibitem{21} M. Aganagic, A. Klemm, and C. Vafa, \emph{\textquotedblleft
Disk Instantons, Mirror Symmetry and the Duality Web,\textquotedblright }
hep-th/0105045.

\bibitem{G1} J.Bryan, R.Pandharipande, \emph{Curves in Calabi-Yau and
topological quantum field theory}, Duke Math.126 (2005) 369-396. J.Bryan,
R.Pandharipande, \emph{On the rigidity of stable maps to Calabi-Yau
threefolds}, Geometry Topology Monographs 8 (2006) 97-104

\bibitem{h} Jim Bryan and Rahul Pandharipande, \emph{The local Gromov-Witten
theory of curves}, math.AG/0411037 v3 2006

\bibitem{G2} D.Karp, C.Liu, M.Marino, \emph{The local Gromov-Witten
invariants of configuration of rational curves}, Geometry Topology
Monographs 10 (2006) 115-168

\bibitem{3} H. Ooguri, A. Strominger, C. Vafa, \emph{Black Hole Attractors
and the Topological String,} Phys. Rev. \textbf{D70} (2004) 106007,
hep-th/0405146.

\bibitem{4} C. Vafa, \emph{\ Two dimensional Yang-Mills, black holes and
topological strings},hep-th/0406058.

\bibitem{5} \ A. Dabholkar, \emph{\ Exact counting of black hole microstates}%
, Phys. Rev. Lett. \textbf{94} (2005) 241-301, hep-th/0409148.

\bibitem{6} H. Ooguri, C. Vafa, E. Verlinde, \emph{Hartle-Hawking
Wave-Function for Flux Compactifications}, Lett. Math. Phys. \textbf{74}
(2005) 311-342, hep-th/0502211.

\bibitem{7} R. Dijkgraaf, R. Gopakumar, H. Ooguri, C. Vafa, \emph{Baby
Universes in String Theory}, Phys. Rev. \textbf{D73} (2006) 066002,
hep-th/0504221.

\bibitem{71} A. Belhaj, L. B. Drissi, E. H. Saidi, A. Segui, $\mathcal{N}=2$
\emph{Supersymmetric Black Attractors in Six and Seven Dimensions},
arXiv:0709.0398, Nuclear Physics B 796 [FS] (2008) 521--580.

\bibitem{72} El Hassan Saidi, Moulay Brahim Sedra, \emph{Topological string
in harmonic space and correlation functions in S\symbol{94}3 stringy
cosmology.}Nucl.Phys.B748:380-457,2006, hep-th/0604204

\bibitem{8} M. Aganagic, A. Neitzke, C. Vafa, \emph{BPS Microstates and the
Open Topological String Wave Function}, hep-th/0504054

\bibitem{80} E. Witten, \emph{\textquotedblleft Topological Sigma
Models,\textquotedblright } Commun. Math. Phys. 118, 411 (1988).

\bibitem{801} M. Bershadsky, S. Cecotti, H. Ooguri, and C. Vafa, \emph{%
\textquotedblleft Holomorphic Anomalies in Topological Field
Theories,\textquotedblright } Nucl. Phys. 405B (1993) 279-304;

\bibitem{81} H. Ooguri and C. Vafa, \emph{\textquotedblleft Knot invariants
and topological strings,\textquotedblright } Nucl. Phys. B 577,419 (2000),
[arXiv:hep-th/9912123].

\bibitem{82} J. Walcher, \emph{\textquotedblleft Extended Holomorphic
Anomaly and Loop Amplitudes in Open Topological, String,\textquotedblright }
arXiv[hep-th]:0705.4098 .

\bibitem{83} S. Yamaguchi and S. T. Yau, \emph{\textquotedblleft Topological
string partition functions as polynomials,\textquotedblright } JHEP 0407,
047 (2004) [arXiv:hep-th/0406078].

\bibitem{830} M. Alim and J. D. Lange, \emph{\textquotedblleft Polynomial
Structure of the (Open) Topological String Partition
Function,\textquotedblright } JHEP 0710, 045 (2007) arXiv [hep-th]:0708.2886.

\bibitem{84} M. Aganagic, R. Dijkgraaf, A. Klemm, M. Marino and C. Vafa,
\emph{\textquotedblleft Topological strings and integrable
hierarchies,\textquotedblright } Commun. Math. Phys. 261, 451 (2006),
[arXiv:hep-th/0312085].

\bibitem{85} V. Bouchard, B. Florea and M. Marino, \emph{\textquotedblleft
Topological open string amplitudes on orientifolds,\textquotedblright } JHEP
0502, 002 (2005) [arXiv:hep-th/0411227].

\bibitem{850} M. Aganagic and C. Vafa, \emph{\textquotedblleft Mirror
Symmetry, D-Branes and Counting Holomorphic Discs,\textquotedblright }
arXiv:hep-th/0012041.

\bibitem{9} M. Aganagic, A. Klemm, M. Marino, C. Vafa, \emph{The Topological
Vertex}, Commun. Math. Phys. \textbf{254} (2005) 425-478, hep-th/0305132.

\bibitem{10} A. Iqbal, A-K Kashani-Poor, \emph{The Vertex on a Strip, }%
hep-th/0410174, \emph{Instanton Counting and Chern-Simons Theory}, Adv.
Theor. Math. Phys. \textbf{7} (2004) 457-497, hep-th/0212279.

\bibitem{11} Amer Iqbal, Can Kozcaz, Cumrun Vafa, \emph{The Refined
Topological Vertex}, arXiv:hep-th/0701156,

\bibitem{12} Piotr Sulkowski, \emph{Crystal Model for the Closed Topological
Vertex Geometry}, JHEP 0612 (2006) 030, arXiv:hep-th/0606055,

\bibitem{12a} A. Iqbal, C. Kozcaz and C. Vafa, \emph{\textquotedblleft The
refined topological vertex,\textquotedblright\ }arXiv:hep-th/0701156.

\bibitem{12b} Masato Taki, Refined Topological Vertex and Instanton Counting
, arXiv::hep-th/0710.1776

\bibitem{120} N. A. Nekrasov, \emph{\textquotedblleft Seiberg-Witten
prepotential from instanton counting,\textquotedblright } Adv. Theor. Math.
Phys. 7, 831 (2004) [arXiv:hep-th/0206161]

\bibitem{1201} N. Nekrasov and A. Okounkov, \emph{\textquotedblleft
Seiberg-Witten theory and random partitions,\textquotedblright\ }%
arXiv:hepth/0306238.

\bibitem{1202} H. Nakajima and K. Yoshioka, \emph{\textquotedblleft
Instanton counting on blowup. I. 4-dimensional pure gauge
theory,\textquotedblright } Invent. Math 162, no. 2, 313 (2005)
[arXiv:math.A.G/0306198].

\bibitem{1203} A. Braverman and P. Etingof, \emph{\textquotedblleft
Instanton counting via affine Lie algebras II: from Whittaker vectors to the
Seiberg-Witten prepotential\textquotedblright } arXiv:math.AG/0409441.

\bibitem{121} A. Iqbal and A. K. Kashani-Poor, \emph{\textquotedblleft
Instanton counting and Chern-Simons theory,\textquotedblright } Adv.
Theor.Math. Phys. 7, 457 (2004) [arXiv:hep-th/0212279]. \textquotedblleft
\emph{SU(N) geometries and topological string amplitudes,\textquotedblright }
Adv. Theor. Math. Phys. 10, 1 (2006) [arXiv:hep-th/0306032].

\bibitem{123} T. Eguchi and H. Kanno, \emph{\textquotedblleft Topological
strings and Nekrasov's formulas,\textquotedblright } JHEP 0312, 006 (2003)
[arXiv:hep-th/0310235].

\bibitem{124} T. J. Hollowood, A. Iqbal and C. Vafa, \emph{\textquotedblleft
Matrix Models, Geometric Engineering and Elliptic Genera,\textquotedblright }
arXiv:hep-th/0310272.

\bibitem{125} C. Vafa, \textquotedblleft Two dimensional Yang-Mills, black
holes and topological strings,\textquotedblright\ hepth/0406058.

\bibitem{13} N. Caporaso, M. Cirafici, L. Griguolo, S. Pasquetti, D.
Seminara, R. J. Szabo, \emph{\ Topological Strings, Two-Dimensional
Yang-Mills Theory and Chern-Simons Theory on Torus Bundles}, hep-th/0609129.

\bibitem{14} R. Ahl Laamara, A. Belhaj, L.B. Drissi, E.H. Saidi, \emph{Black
Holes in Type IIA String on Calabi-Yau Threefolds with Affine ADE Geometries
and q-Deformed 2d Quiver Gauge Theories, }arXiv:hep-th/0611289,
Nucl.Phys.B776:287-326,2007

\bibitem{15} S. Katz, P. Mayr, C. Vafa, \emph{Mirror symmetry and Exact
Solution of 4D N=2 Gauge Theories I}, Adv. Theor. Math. Phys. \textbf{1}
(1998) 53-114, hep-th/9706110.

\bibitem{16} M. Ait Ben Haddou, A. Belhaj, E.H. Saidi, \emph{Geometric
Engineering of N=2 CFT}$_{4}$\emph{s based on Indefinite Singularities:
Hyperbolic Case,} Nucl. Phys. \textbf{B674} (2003) 593-614, hep-th/0307244.

\bibitem{17} R. Ahl Laamara, M. Ait Ben Haddou, A. Belhaj, L.B. Drissi, E.H.
Saidi, \emph{RG Cascades in Hyperbolic Quiver Gauge Theories}, Nucl. Phys.
\textbf{B702} (2004) 163-188, hep-th/0405222.

\bibitem{18} Edward Witten, \emph{Phases of N=2 Theories In Two Dimensions},
Nucl.Phys. B403 (1993) 159-222, arXiv:hep-th/9301042

\bibitem{19} N.C. Leung, C. Vafa, \emph{Branes and Toric Geometry}, Adv.
Theor. Math.Phys. \textbf{2} (1998) 91-118, hep-th/9711013.

\bibitem{20} L.B. Drissi, H. Jehjouh, E.H. Saidi, \emph{Topological String} {
on Local Elliptic Curve}\\{ with Large Complex Structure}, Afr
Journal Of Mathematical Physics, Volume 6 (2007) 83-91.

\bibitem{DJS} Lalla Btissam Drissi, Houda Jehjouh, El Hassan Saidi, \emph{
Generalized MacMahon G(q) as q-deformed CFT Correlation Function},
arXiv:0801.2661, To appear in Nucl Phys B (2008),

\bibitem{ge} S. Katz, P. Mayr, C. Vafa, \emph{Mirror symmetry and Exact
Solution of 4D N=2 Gauge Theories} I, Adv.Theor.Math.Phys. 1 (1998) 53-114,
arXiv:hep-th/9706110

\bibitem{gf} A. Belhaj, A. Elfallah, E.H. Saidi, \emph{On the non-simply
laced mirror geometries in type II strings},
Class.Quant.Grav.17:515-532,2000, \emph{On the Affine D(4) mirror geometry},
Class.Quant.Grav.16:3297-3306,1999
\end{thebibliography}
\end{document}